\documentclass[11pt,a4paper,final]{article}

\usepackage[]{lineno}

\usepackage[a4paper,
		right=2cm,
		left=4cm,
		top=2.5cm,
		bottom=2cm,
		textwidth=183mm,
		includefoot,
		foot=1cm,
		head=2cm]{geometry}
		
\usepackage{mdframed}
\usepackage{calc}
\newlength{\mylength}
\newlength{\figurewidth}
\usepackage{amsmath,amsfonts, amssymb, amsthm}

\usepackage{wasysym}
\usepackage{xcolor, framed}
\definecolor{shadecolor}{rgb}{0.5, 0.5, 0.5}

\usepackage{sectsty}
\sectionfont{\color{shadecolor}}
\subsectionfont{\normalsize}

\usepackage[doublespacing]{setspace}

\usepackage[]{graphicx}

\usepackage{multicol}
\usepackage{multirow}

\usepackage{sidecap}
\sidecaptionvpos{figure}{c}
\usepackage[font=small,labelfont=bf,sc]{caption}
\usepackage{wrapfig}

\usepackage{siunitx}

\usepackage{enumerate, listings}

\newcommand{\E}[2][]{\ensuremath{\langle #2 \rangle}_{#1}} 
\newcommand{\Var}[1]{\ensuremath{\mathrm{Var}[ #1 ]} }
\newcommand{\Prob}[1]{\ensuremath{\mathrm{P}[ #1 ]}} 
\newcommand{\cond}{\ensuremath{\,| \,}}

\newcommand{\sh}{\ensuremath{\hat m}}
\newcommand{\mh}{\ensuremath{\hat m}}

\newcommand{\invivo}{\textit{in vivo}}

\newsavebox\myboxA
\newsavebox\myboxB
\newlength\mylenA
\newcommand*\widebar[2][0.8]{%
    \sbox{\myboxA}{$\m@th#2$}%
    \setbox\myboxB\null
    \ht\myboxB=\ht\myboxA%
    \dp\myboxB=\dp\myboxA%
    \wd\myboxB=#1\wd\myboxA
    \sbox\myboxB{$\m@th\overline{\copy\myboxB}$}
    \setlength\mylenA{\the\wd\myboxA}
    \addtolength\mylenA{-\the\wd\myboxB}%
    \ifdim\wd\myboxB<\wd\myboxA%
       \rlap{\hskip 0.5\mylenA\usebox\myboxB}{\usebox\myboxA}%
    \else
        \hskip -0.5\mylenA\rlap{\usebox\myboxA}{\hskip 0.5\mylenA\usebox\myboxB}%
    \fi}

\theoremstyle{definition}

\theoremstyle{remark}

\newcommand{\change}[1]{\textcolor{black}{#1}}

\renewcommand{\thesubsection}{\arabic{subsection}}

\newcommand{\beginsupplement}{%
        \setcounter{table}{0}
        \renewcommand{\thetable}{S\arabic{table}}%
        \setcounter{figure}{0}
        \renewcommand{\thefigure}{S\arabic{figure}}%
        \setcounter{equation}{0}
        \renewcommand{\theequation}{S\arabic{equation}}%
        \setcounter{theorem}{0}
        \renewcommand{\thetheorem}{S\arabic{theorem}}%
        \setcounter{definition}{0}
        \renewcommand{\thedefinition}{S\arabic{definition}}%
        \renewcommand{\thesubsection}{Supp. \arabic{subsection}}
     }

\usepackage[authoryear,comma]{natbib}
\newcommand{\autocite}{\citep}

\author{J. Wilting$^1$ and V. Priesemann$^{1,2}$ \\ {\small $^1$Max-Planck-Institute for Dynamics and Self-Organization, G\"ottingen} \\ {\small $^2$Bernstein-Center for Computational Neuroscience, G\"ottingen}}
\date{ \normalsize Preprint, \today}

\begin{document}
   
      {\Huge  \noindent 
      \change{Between perfectly critical and fully irregular: a reverberating model captures and predicts cortical spike propagation}}\\
   
   {\large \noindent J. Wilting$^1$ \& V. Priesemann$^{1,2,*}$ }\\
   
   \noindent
   $^1$Max-Planck-Institute for Dynamics and Self-Organization, Am Fa\ss berg 17, 37077  G\"ottingen, Germany; $^2$Bernstein-Center for Computational Neuroscience, G\"ottingen, Germany\\
   $^*$ viola.priesemann@ds.mpg.de\\   
   \textbf{Keywords: balanced state, criticality, perturbations, timescales}
       
\vspace*{0.5cm}

\noindent
\textbf{
\noindent
\change{
Knowledge about the collective dynamics of cortical spiking is very informative about the underlying coding principles.
However, even most basic properties are not known with certainty, because their assessment is hampered by spatial subsampling, i.e. the limitation that only a tiny fraction of all neurons can be recorded simultaneously with millisecond precision.
Building on a novel, subsampling-invariant estimator, we fit and carefully validate a minimal model for cortical spike propagation. The model interpolates between two prominent states: asynchronous and critical.
We find neither of them in cortical spike recordings across various species, but instead identify a narrow ``reverberating'' regime.
This approach enables us to predict yet unknown properties from very short recordings  and for every circuit individually, including responses to minimal perturbations, intrinsic network timescales, and the strength of external input compared to recurrent activation – thereby informing about the underlying coding principles for each circuit, area, state and task.
}
}

\vspace*{0.5cm}





\section*{Introduction}


\change{
In order to understand how each cortical circuit or network processes its input, it would be desirable to first know its basic dynamical properties.
For example, knowing which impact one additional spike has on the network \autocite{London2010} would give insight into the amplification of small stimuli \autocite{Douglas1995,Suarez1995,Miller2016}.
Knowing how much of cortical activity can be attributed to external activation or internal activation \autocite{Reinhold2015} would allow to gauge how much of cortical activity is actually induced by stimuli, or rather internally generated, for example in the context of predictive coding \autocite{Rao1999,Clark2013}.
Knowing the intrinsic network timescale \autocite{Murray2014a} would inform how long stimuli are maintained in the activity and can be read out for short term memory \autocite{Buonomano1995,Wang2002,Jaeger2007,Lim2013}.
However, not even these basic properties of cortical network dynamics are generally known with certainty.
}

\change{
In the past, insights about these network properties have been strongly hampered by the inevitable limitations of spatial subsampling, i.e. the fact that only a tiny fraction of all neurons can be recorded experimentally with millisecond precision.
Such spatial subsampling fundamentally limits virtually any recording and hinders inferences about the collective response of cortical networks \autocite{Priesemann2009,Ribeiro2010,Priesemann2014,Ribeiro2014,Levina2017}.
}

\change{To describe network responses, two contradicting hypotheses have competed for more than a decade, and are the subjects of ongoing scientific debate}:
One hypothesis suggests that collective dynamics are ``asynchronous-irregular'' (AI) \autocite{Burns1976, Softky1993,Stein2005}, i.e. neurons spike independently of each other and in a Poisson manner, which may reflect a balanced state \autocite{Vreeswijk1996a,Brunel2000}.
The other hypothesis proposes that neuronal networks operate at criticality \autocite{Beggs2003,Levina2007,Levina2009a,Munoz2018,Beggs2012,Plenz2014,Tkacik2014,Humplik2017}. \change{Criticality is a particular state at a phase transition, characterized by high sensitivity and long-range correlations in space and time}.

These hypotheses have distinct implications for the coding strategy of the brain.
The typical balanced state minimizes redundancy \autocite{Barlow1961,Atick1992,Bell1997,VanHateren1998,Hyvarinen2000}, supports fast network responses \autocite{Vreeswijk1996a}, and shows vanishing autocorrelation time \change{or network timescale}.
In contrast, criticality in models optimizes performance in tasks that profit from extended reverberations of activity in the network \autocite{Bertschinger2004,Haldeman2005,Kinouchi2006,Wang2011b,Boedecker2012,Shew2013,DelPapa2017}.

Surprisingly, there is experimental evidence for both AI and critical states in cortical networks, although both states are clearly distinct.
Evidence for the AI state is based on characteristics of single neuron spiking, resembling a Poisson process, i.e. exponential inter spike interval (ISI) distributions and a Fano factor $F$ close to unity \change{\autocite{Burns1976,Tolhurst1981,Vogels1989,Softky1993,Gur1997,Steveninck1997,Kara2000,Carandini2004}}. 
Moreover, spike count cross-correlations \autocite{Ecker2010a,Cohen2011} are small.
In contrast, evidence for criticality was typically obtained from a population perspective instead, and assessed neuronal avalanches, i.e. spatio-temporal clusters of activity \autocite{Beggs2003, Pasquale2008, Priesemann2009, Friedman2012, Tagliazucchi2012, Shriki2013}, whose sizes are expected to be power-law distributed if networks are critical \autocite{Bak1987}. 
Deviations from power-laws, typically observed for spiking activity in awake animals \autocite{Bedard2006, Hahn2010, Ribeiro2010,Priesemann2014},  were attributed to subsampling effects \autocite{Priesemann2009,Ribeiro2010,Priesemann2013,Girardi-Schappo2013,Priesemann2014,Ribeiro2014,Levina2017}. 
Hence, different analysis approaches provided evidence for one or the other hypothesis about cortical dynamics.

We here resolve the contradictory results about cortical dynamics,  building on a subsampling-invariant approach presented in a companion study \autocite{Wilting2018}: 
(i) we establish an analytically tractable minimal model for \textit{in vivo}-like activity, which can interpolate from AI to critical dynamics \change{(Fig. \ref{fig:states}\textbf{a})};
(ii) we estimate the dynamical state of cortical activity based on a novel, subsampling-invariant estimator \autocite{Wilting2018} \change{(Figs. \ref{fig:states}\textbf{b -- d})};
(iii) the model reproduces a number of dynamical properties of the network, which are experimentally accessible and enable us to validate our approach;
(iv) we predict a number of yet unknown network properties, including the expected number of spikes triggered by one additional spike, the intrinsic network timescale, the distribution of the total number of spikes triggered by a single extra action potential, and the fraction of activation that can be attributed to afferent external input compared to recurrent activation in a cortical network.

\section*{Material and Methods}


We analyzed \textit{in vivo} spiking activity from Macaque monkey prefrontal cortex during a short term memory task \autocite{Pipa2009}, anesthetized cat visual cortex with no stimulus \autocite{Blanche2006,Blanche2009}, 
and rat hippocampus during a foraging task \autocite{Mizuseki2009a,Mizuseki2009} (\ref{sec:supp_experiments}).
We compared the recordings of each experimental session to results of a minimal model of  spike propagation, which is detailed in the following.

\subsubsection*{Minimal model of spike propagation}

\change{
To gain an intuitive understanding of our mathematical approach, make a thought experiment in your favorite spiking network: apply one additional spike to an excitatory neuron, in analogy to the approach by \cite{London2010}.
How does the network respond to that perturbation?
As a first order approximation, one quantifies the number of spikes that are directly triggered \textit{additionally} in all postsynaptic neurons.
This number may vary from trial to trial, depending on the membrane potential of the postsynaptic neurons.
However, what interests us most is $m$, the \textit{mean number of spikes triggered by the one extra spike}.
Any of these triggered spikes can in turn trigger spikes in their postsynaptic neurons in a similar manner, and thereby the perturbation may cascade through the system.
}

\change{
In the next step, assume that perturbations are started continuously at rate $h$, for example through afferent input from other brain areas or sensory modalities.
Together, this leads to the mathematical framework of a branching model \autocite{Harris1963, Heathcote1965, Pakes1971, Beggs2003, Haldeman2005, Ribeiro2010, Priesemann2013, Priesemann2014}.
This framework describes the number of active neurons $A_t$ in discrete time bins of length $\Delta t$.
Here, $\Delta t$ should reflect the propagation time of spikes between neurons.
Formally, each spike $i$ at the time bin $t$ excites a random number $Y_{t,i}$ of postsynaptic spikes, on average $m = \E{Y_{t,i}}$.
The activity $A_{t+1}$, i.e. the total number of spikes in the next time bin is then defined as the sum of the postsynaptic spikes of all current spikes $A_t$, as well as the input $h_t$:
}

\begin{equation}
    A_{t+1} = \sum_{i=1}^{A_t} Y_{t,i} + h_t.
    \label{eq:simple_bp}
\end{equation}

\noindent
\change{
This generic spiking model can generate dynamics spanning AI and critical states depending on the input \autocite{Zierenberg2018}, and hence is well suited to probe network dynamics \textit{in vivo} (see \ref{sec:supp_bps} for details).
Most importantly, this framework enables us to infer $m$ and other properties from the ongoing activity proper.
Mathematical approaches to infer $m$ are long known if the full network is sampled \autocite{Heyde1972,Wei1991}.
Under subsampling, however, it is the novel estimator described in \cite{Wilting2018} that for the first time allows an unbiased inference of $m$, even if only a tiny fraction of neurons is sampled. 
}

\change{
A precise estimate of $m$ is essential, because the dynamics of the model is mainly governed by $m$ (Fig. \ref{fig:states}\textbf{a}).
Therefore, after inferring $m$, a number of quantities can be analytically derived, and others can be obtained by simulating a branching model, which is constrained by the experimentally measured $m$ and the spike rate.
}

\subsubsection*{Simulation}
\change{We simulated a branching model by mapping a branching process (Eq. \eqref{eq:simple_bp} and \ref{sec:supp_bps}) onto a random network of $N=10,000$ neurons in the annealed disorder limit} \autocite{Haldeman2005}. 
An active neuron activated each of its \change{$\kappa = 4$} postsynaptic neurons with probability $p = m / \change{\kappa}$.
Here, the activated postsynaptic neurons were drawn randomly without replacement at each step, thereby avoiding that two different active neurons would both activate the same target neuron.
The branching model is critical for $m = 1$ in the infinite-size limit, and subcritical (supercritical) for $m < 1$ ($m > 1$).
We modeled input to the network at rate $h$ by Poisson activation of each neuron at rate $h/N$.
Subsampling \autocite{Priesemann2009} was applied to the model by sampling the activity of $n$ neurons only, which were selected randomly before the simulation, and neglecting the activity of all other neurons.
Thereby, instead of the full activity $A_t$, only the subsampled activity $a_t$ was considered for observation.

If not stated otherwise, simulations were run for $L=10^7$ time steps (corresponding to $\sim$\SI{11}{h}). Confidence intervals were estimated according to \cite{Wilting2018} from $B=100$ realizations of the model, both for simulation and experiments.

\change{
We compared the experimental recordings to three different models: AI, near-critical, and reverberating.
All three models were set up to match the experiment in the number of sampled neurons $n$ and firing rate $R=\E{a_t} / (n \cdot \Delta t)$.
The AI and near-critical models were set up with branching ratios of $m=0$ or $m=0.9999$, respectively.
In addition, the reverberating model matched the recording in $m = \mh$, where $\mh$ was estimated from the recording using the novel subsampling-invariant estimator (see below).
For all models, we chose a full network size of $N=10^4$ and then determined the appropriate input $h = R \, \Delta t \, N \, (1 - m)$ in order to match the experimental firing rate.
Exemplarily for the cat recording, which happened to represent the median $\mh$, this yielded $\mh = 0.98$,  $n=50$, and $R=\SI{7.25}{Hz}$.
From these numbers, $h=290$, $h=5.8$ and $h=0.029$ followed for the AI, reverberating, and near-critical models, respectively.
}

\begin{figure*}[h!]
\includegraphics[width=\textwidth]{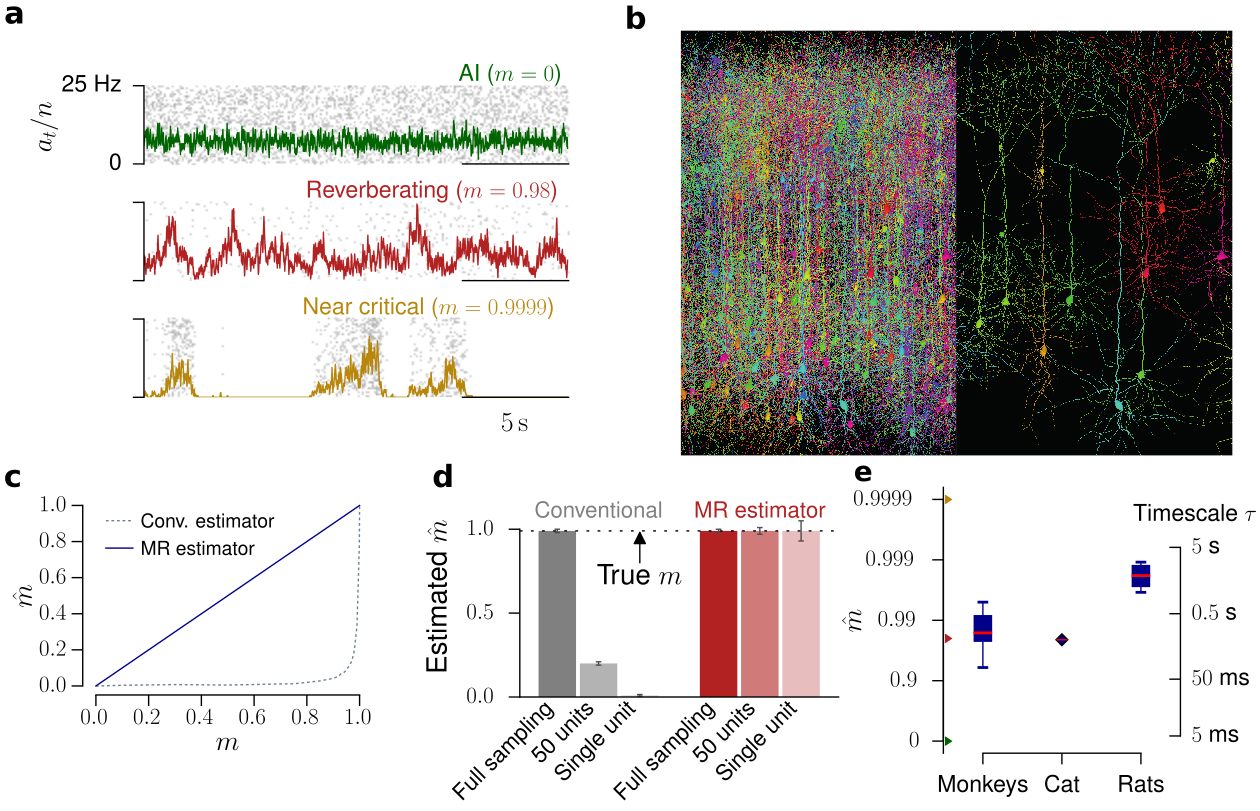}
\caption{\textbf{Reverberating versus critical and irregular dynamics under subsampling.}
\textbf{a}. Raster plot and population rate $a_t$ for networks with different spike propagation parameters or neural efficacy $m$.
They exhibit vastly different dynamics, which readily manifest in the population activity.
\textbf{b}. When recording spiking activity, only a small subset of all neurons can be sampled with millisecond precision. This spatial subsampling can hinder correct inference of collective properties of the whole network; figure created using TREES \autocite{Cuntz2010} and reproduced from \cite{Wilting2018}. 
\textbf{c}. Estimated branching ratio $\mh$ as a function of the simulated, true branching ratio $m$, inferred from subsampled activity (100 out of 10,000 neurons). While the conventional estimator misclassified $m$ from this subsampled observation (gray, dotted line), the novel  multistep regression (MR) estimator returned the correct values 
\textbf{d}. For a reverberating branching model with $m=0.98$, the conventional estimator inferred $\mh = 0.21$ or $\mh = 0.002$ when sampling 50 or 1 units respectively, in contrast to MR estimation, which returned the correct $\mh$ even under strong subsampling.
\textbf{e}. Using the novel MR estimator, cortical network dynamics in monkey prefrontal cortex, cat visual cortex, and rat hippocampus consistently showed reverberating dynamics, with $0.94 <\mh < 0.991$ (median $\mh = 0.98$ over all experimental sessions, boxplots indicate median / 25\% -- 75\% / 0\% -- 100\% over experimental sessions per species). These correspond to intrinsic network timescales between \SI{80}{ms} and \SI{2}{s}.
}
\label{fig:states}
\end{figure*}

In Fig. \ref{fig:validation}, the reverberating branching model was also matched to the length of the cat recording of \SI{295}{s}. To test for stationarity, the cat recording and the reverberating branching model were split into 59 windows of \SI{5}{s} each, before estimating $m$ for each window.
In Fig. \ref{fig:states}\textbf{c}, subcritical and critical branching models with $N=10^4$ and $\E{A_{t}} = 100$ were simulated, and $n=100$ units sampled.


\change{
\subsubsection*{Subsampling-invariant estimation of $\mh$}
}

\change{
Details on the analysis are found in \ref{sec:supp_analysis}. For each experimental recording, we collected the spike times of all recorded units (both single and multi units) into one single train of population spike counts $a_{t}$, where $a_{t}$ denotes how many neurons spiked in the $t^{th}$ time bin $\Delta t$.
If not indicated otherwise, we used $\Delta t = \SI{4}{ms}$, reflecting the propagation time of spikes from one neuron to the next.
}

\change{
From these experimental time series, we estimated $\hat{m}$ using the multistep regression (MR) estimator described in all detail in \cite{Wilting2018}.
In brief, we calculated the linear regression slope $r_{k\, \Delta t}$, which describes the linear statistical dependence of $a_{t+k}$ upon $a_t$, for different time lags $\delta t = k \Delta t$ with $k=1,\ldots,k_\mathrm{max}$.
In our branching model, these slopes are expected to follow the relation $r_{\delta t} = b \cdot \mh^{\delta t / \Delta t}$ (or $r_{k\, \Delta t} = b \cdot \mh^k$), where $b$ is an unknown parameter that depends on the higher moments of the underlying process and the degree of subsampling.
However, it can be partialled out, allowing for an estimation of $m$ without further knowledge about $b$.
Throughout this study we chose $k_\mathrm{max} = 2500$ (corresponding to \SI{10}{s}) for the rat recordings, $k_\mathrm{max} = 150$ (\SI{600}{ms}) for the cat recording, and $k_\mathrm{max} = 500$ (\SI{2000}{ms}) for the monkey recordings, assuring that $k_\mathrm{max} \, \Delta t$ was always in the order of multiple intrinsic network timescales. 
In order to test for the applicability of a MR estimation, we used a set of conservative tests \autocite{Wilting2018}.
The exponential relation can be rewritten as an exponential autocorrelation function $r_{\delta t} = b \, m^{\delta t / \Delta t} = \exp( \ln m \, \delta t / \Delta t ) = \exp (-\delta t / \tau )$, where the intrinsic network timescale $\tau$ relates to $m$ as $m = \exp ( - \Delta  t / \tau )$.
While the precise value of $m$ depends on the choice of the bin size $\Delta t$ and should only be interpreted together with the bin size ($\Delta t = \SI{4}{ms}$ throughout this study), the intrinsic network timescale is independent of $\Delta t$.
Therefore, we report both values in the following.
}

\section*{Results}

\subsection*{Reverberating spiking activity \textit{in vivo}}
We applied MR estimation to the binned population spike counts $a_t$ of the recorded neurons of each experimental session across three different species (see methods).
We identified a limited range of branching ratios \textit{in vivo}:
in the experiments $\mh$ ranged from $0.963$ to $0.998$ (median $\mh=0.98$\change{, for a bin size of $\Delta t = \SI{4}{ms}$}), which is only a narrow window in the continuum from AI ($m=0$) to critical ($m=1$).
\change{
    For these values of $\mh$ found in cortical networks, the corresponding  $\tau$ 
    are between \SI{100}{ms} and \SI{2}{s} (median \SI{247}{ms}, Figs. \ref{fig:states}\textbf{e}, \ref{fig:supp_animal_data}).
   }
This clearly suggests that spiking activity \textit{in vivo} is neither AI-like, nor consistent with a critical state.
Instead, it is poised in a regime that, unlike critical or AI, does not maximize one particular property alone but may flexibly combine features of both \autocite{Wilting2018a}.
Without a prominent characterizing feature, we name it the \textit{reverberating regime}, stressing that activity reverberates (different from the AI state) at timescales of hundreds of milliseconds (different from a critical state, where they can persist infinitely).


\begin{figure*}[h!]
\centering
\includegraphics[width=\textwidth]{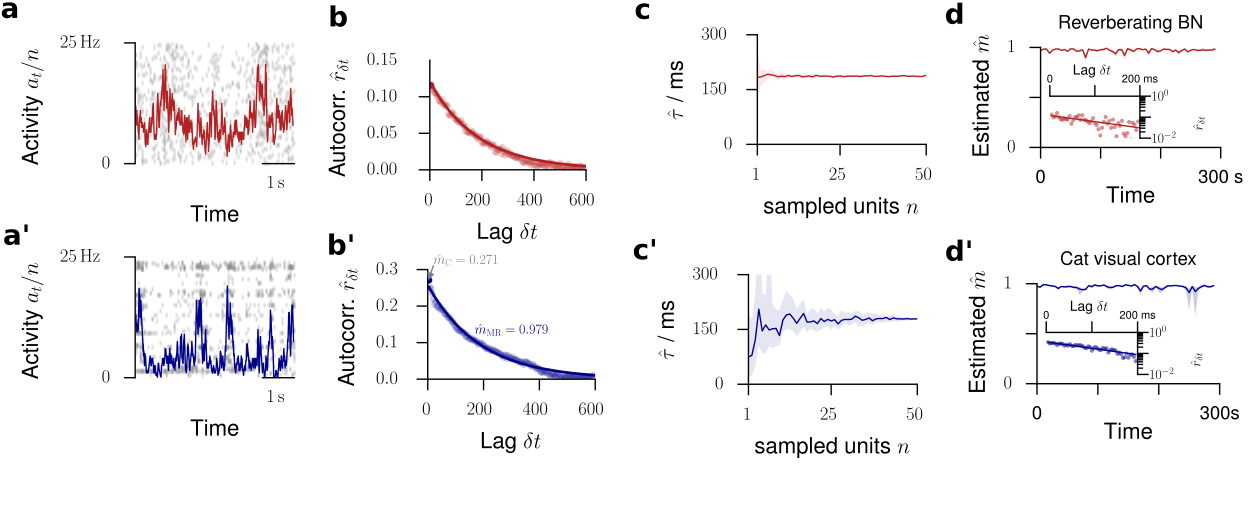}
\caption{\textbf{Validation of the model assumptions.}
The top row displays properties from a  reverberating model, the bottom row spike recordings from cat visual cortex.
\textbf{a/a'}. Raster plot and population activity $a_t$ within bins of $\Delta t =\SI{4}{ms}$, sampled from $n=50$ neurons.
\textbf{b/b'}. Multistep regression (MR) estimation from the subsampled activity (\SI{5}{min} recording). The predicted exponential relation $r_{\delta t} \sim m^{\delta t / \Delta t} = \exp ( - \delta t / \tau )$ provides a validation of the applicability of the model.  
The experimental data are fitted by this exponential with remarkable precision.
    \change{\textbf{c/c'}.} When subsampling even further, MR estimation always returns the correct timescale $\hat{\tau}$ (or $\mh$) in the model. In the experiment, this invariance to subsampling also holds, down to $n\approx 10$ neurons (shaded area: 16\% to 84\% confidence intervals estimated from 50 subsets of $n$ neurons).
    \change{    \textbf{d/d'}.} The estimated branching parameter $\sh$ for 59 windows of $\SI{5}{s}$ length suggests stationarity of $m$ over the entire recording (shaded area: 16\% to 84\% confidence intervals). The variability in $\mh$ over consecutive windows was comparable for experimental recording and the matched model ($p=0.09$, Levene test).
Insets: Exponential decay exemplified for one example window each.
}
\label{fig:validation}
\end{figure*}

\subsection*{Validity of the approach}
There is a straight-forward verification of the validity of our phenomenological model: it predicts an exponential autocorrelation function $r_{\delta t}$ for the population activity $a_t$.
We found that the activity in cat visual cortex (Figs. \ref{fig:validation}\textbf{a},\textbf{a'}) is surprisingly well described by this exponential fit (Fig. \ref{fig:validation}\textbf{b},\textbf{b'}).
This validation holds to the majority of experiments investigated (14 out of 21, Fig. \ref{fig:supp_animal_data}).

A second verification of our approach is based on its expected invariance under subsampling:
We further subsampled the activity in cat visual cortex by only taking into account spikes recorded from a subset $n'$ out of all available $n$ single units.
As predicted (Fig. \ref{fig:validation}\change{\textbf{c})}, the estimates of $\mh$, or equivalently \change{of the intrinsic network timescale $\hat{\tau}$}, coincided for any subset of single units if at least about five of the available 50 single units were evaluated (Fig. \ref{fig:validation}\change{\textbf{c'}}). Deviations when evaluating only a small subset of units most likely  reflect the heterogeneity within cortical networks.
Together, these results demonstrate that our approach returns consistent results when evaluating the activity of $n \geq 5$ neurons, which were available for all investigated experiments.

\subsection*{Origin of the activity fluctuations}
The fluctuations found in cortical spiking activity, instead of being intrinsically generated, could in principle arise from \change{non-stationary input}, which could in turn lead to misestimation of $m$ \autocite{Priesemann2018}.
This is unlikely for three reasons: First, the majority of experiments passed a set of conservative tests that reject recordings that show any signature of common non-stationarities, as defined in \cite{Wilting2018}.
Second, recordings in cat visual cortex were acquired in absence of any stimulation, excluding stimulus-related non-stationarities.
Third, when splitting the spike recording into short windows, the window-to-window variation of $\sh$ in the recording did not differ from that of stationary \textit{in vivo}-like reverberating models ($p = 0.3$, Figs. \ref{fig:validation}\change{\textbf{d},\textbf{d'}}).
For these reasons the observed fluctuations in the estimates likely originate from the characteristic fluctuations of collective network dynamics within the reverberating regime.

 \begin{figure*}[h!]
 \includegraphics[width=\textwidth]{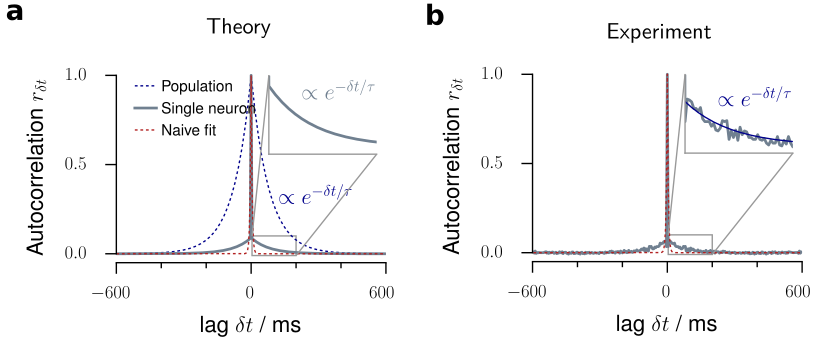}
\caption{
\textbf{MR estimation and intrinsic network timescales.}
\textbf{a}. In a branching model, the autocorrelation function of the population activity decays exponentially with an intrinsic network timescale $\tau$ (blue dotted line). In contrast, the autocorrelation function for single neurons shows a sharp drop from $r_0 = 1$ at lag  $\delta t = 0$ to the next lag $r_{\pm \Delta t}$ (gray solid line).
We showed previously that this drop is a subsampling-induced bias. When ignoring the zero-lag value, the autocorrelation strength is decreased, but the exponential decay and even the value of the intrinsic network timescale $\tau$ of the network activity are preserved (inset). The red, dashed line shows a potential, naive exponential function, fitted to the single neuron autocorrelation function (gray). This naive fit would return a much smaller $\tau$.
\textbf{b}. The autocorrelation function of single neuron activity recorded in cat visual cortex (gray) precisely resembles this theoretical prediction, namely a sharp drop and then an exponential decay (blue, inset), which persists over more than \SI{100}{ms}. A naive exponential fit (red) to the single neuron data would return an extremely short $\tau$.
}
\label{fig:autocorr}
\end{figure*}

\subsection*{Timescales of the network and single units}
The dynamical state described by $m$ directly relates to an exponential autocorrelation function with an intrinsic network timescale $\tau = - \Delta t / \ln m$.
Exemplarily for the cat recording, $m = 0.98$ implies an intrinsic network timescale of $\tau = \SI{188}{ms}$, with $\Delta t = \SI{4}{ms}$ reflecting the spike propagation time from one neuron to the next.
\change{
While the autocorrelation function of the full network activity is expected to show an exponential decay (Fig. \ref{fig:autocorr}\textbf{a}, blue), this is different for the autocorrelation of single neurons -- the most extreme case of subsampling.
We showed that subsampling can strongly decrease the absolute values of the autocorrelation function for any non-zero time lag (Fig. \ref{fig:autocorr}\textbf{a}, gray).
This effect is typically interpreted as a lack of memory, because the autocorrelation of single neurons decays at the order of the bin size (Fig. \ref{fig:autocorr}\textbf{a}, red).
However, ignoring the value at $\delta t = 0$, the floor of the autocorrelation function still unveils the exponential relation.
}
Remarkably, the autocorrelation function of single units in cat visual cortex displayed precisely the shape predicted under subsampling (compare Figs. \ref{fig:autocorr}\textbf{a} and \textbf{b}).

\subsection*{Established methods are biased to identifying AI dynamics}
On the population level, networks with different $m$ are clearly distinguishable (Fig. \ref{fig:states}\textbf{a}).
Surprisingly, single neuron statistics, namely interspike interval (ISI) distributions, Fano factors, conventional estimation of $m$, and the autocorrelation strength $r_{\delta t}$, all returned signatures of AI activity regardless of the underlying network dynamics, and hence these single-neuron properties don't serve as a reliable indicator for the network's dynamical state.

\begin{figure*}[h!]
\includegraphics[width=\textwidth]{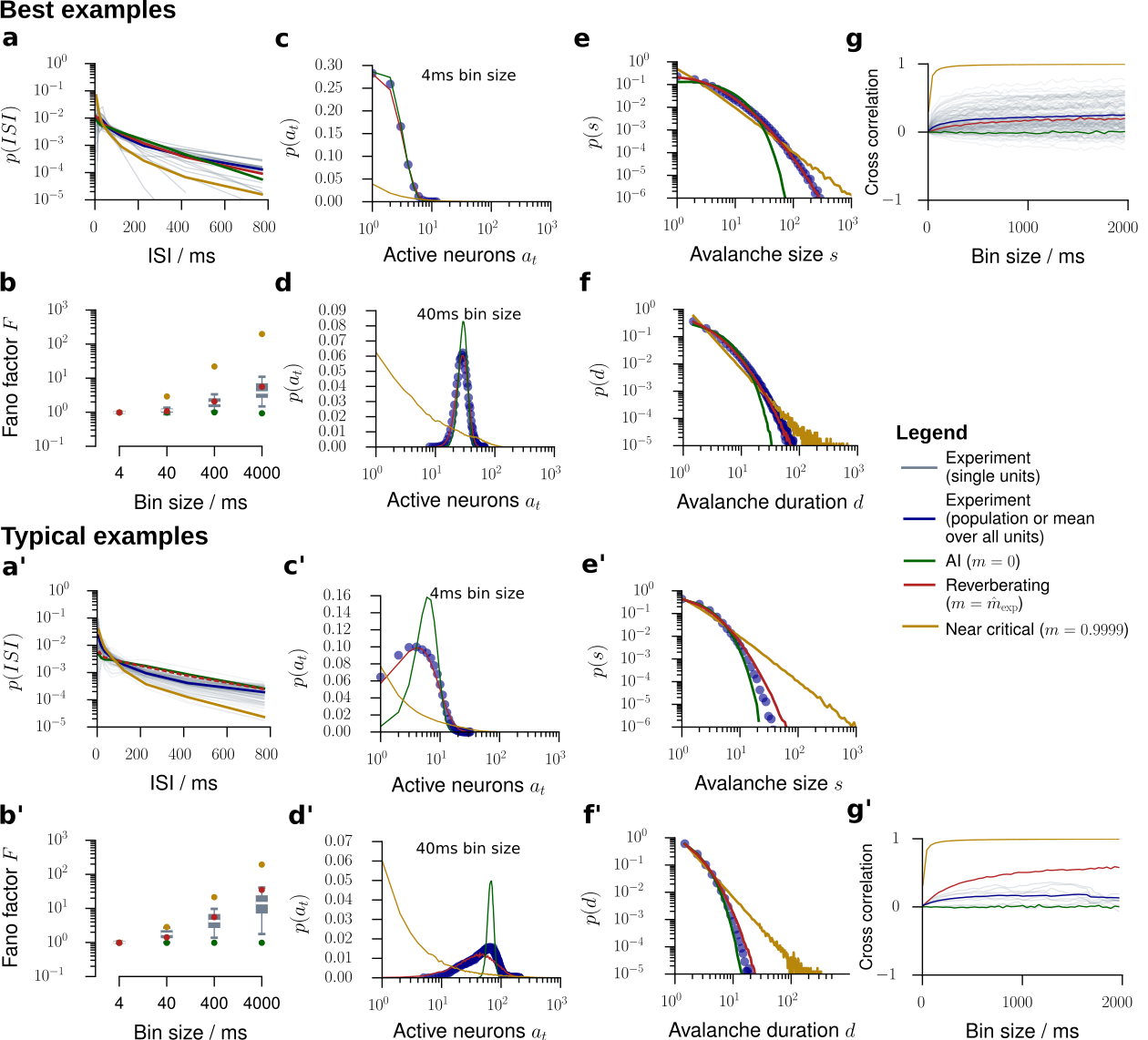}
\caption{\textbf{Model validation for \textit{in vivo} spiking activity.}
We validated our model by comparing experimental results to predictions obtained from the \textit{in vivo}-like, reverberating model, which was matched to the recording in the mean rate, inferred $m$, and number of recorded neurons.
In general, the experimental results  (gray or blue) were best matched by this reverberating model (red), compared to asynchronous-irregular (AI, green) and near-critical (yellow) models.
From all experimental sessions, best examples (top) and typical examples (bottom) are displayed.
For results from all experimental sessions see Figs. \ref{fig:supp_ISIs} to \ref{fig:supp_avalanche_durations}.
\textbf{a}/\textbf{a'}. Inter-spike-interval (ISI) distributions. 
\textbf{b}/\textbf{b'}. Fano factors of single neurons for bin sizes between \SI{4}{ms} and \SI{4}{s}. 
\textbf{c}/\textbf{c'}. Distribution of spikes per bin $p(a_t)$ at a bin size of \SI{4}{ms}.
\textbf{d}/\textbf{d'}. Same as \textbf{c}/\textbf{c'} with a bin size of \SI{40}{ms}.
\textbf{e}/\textbf{e'}. Avalanche size distributions $p(s)$ for all sampled units. AI activity lacks large avalanches, near critical activity produces power-law distributed avalanches, even under subsampling.
\textbf{f}/\textbf{f'}. Same as \textbf{e}/\textbf{e'}, but for the avalanche duration distributions $p(d)$.
\textbf{g}/\textbf{g'}. Spike count cross-correlations ($r_\mathrm{sc}$) as a function of the bin size.
}
\label{fig:animals}
\end{figure*}

\begin{figure}
 \centering
\includegraphics[width=\textwidth]{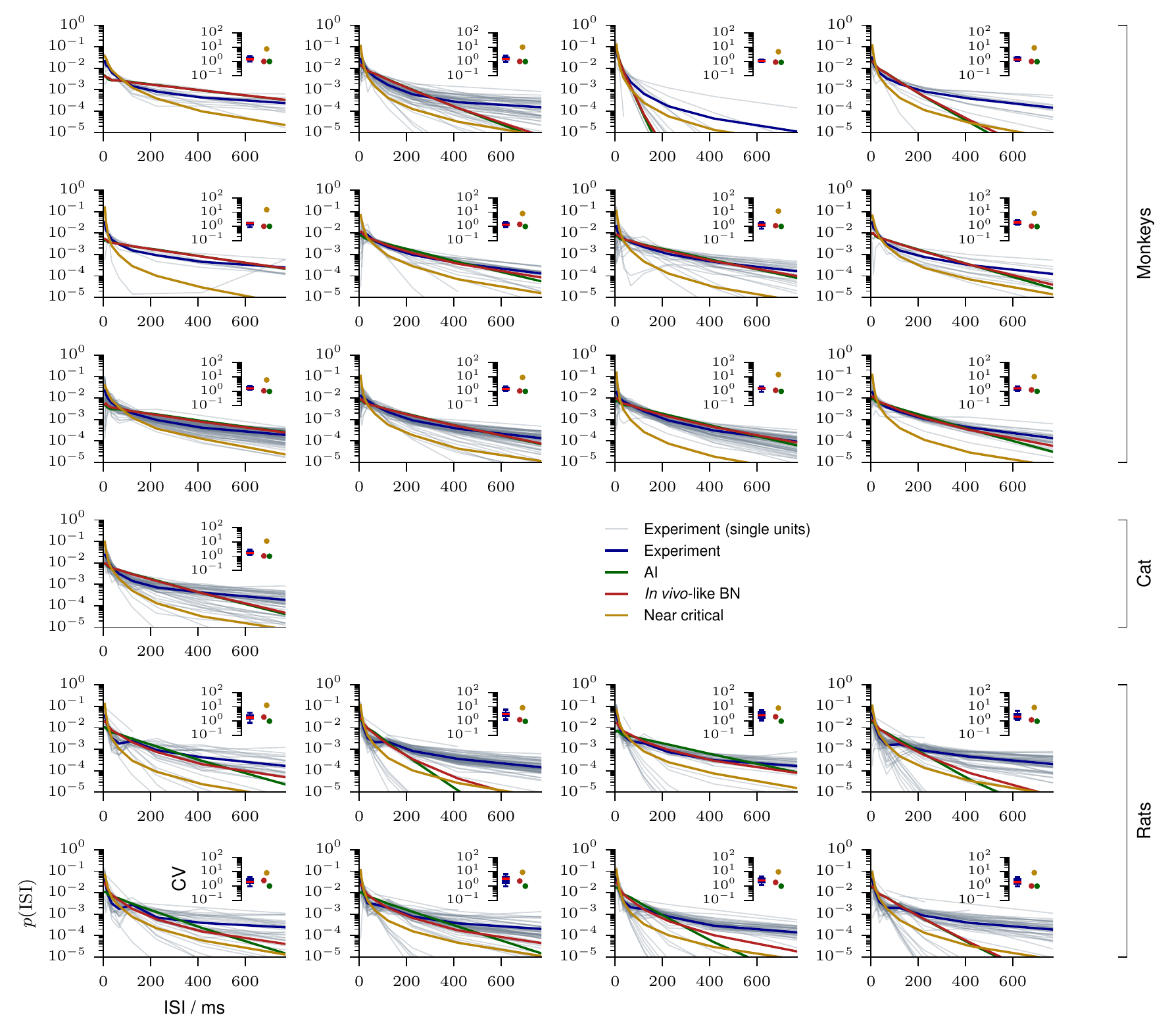}
\caption{\textbf{Interspike interval distribution for individual recording sessions.} Interspike interval (ISI) distributions are shown for individual units of each recording (gray), for the average over units of each recording (blue), as well as for the matched models, either AI (green), \textit{in vivo}-like (red), or near critical (yellow). The insets show the corresponding coefficients of variation (CV). For every experiment AI and \textit{in vivo}-like models are virtually indistinguishable by the ISI distributions.}
\label{fig:supp_ISIs}
\end{figure}

First, exponential interspike interval (ISI) distributions are considered a strong indicator of Poisson-like firing.
Surprisingly, the ISIs of single neurons in the \invivo-like branching model closely followed exponential distributions as well. The ISI distributions were almost indistinguishable for reverberating and AI models (Figs. \ref{fig:animals}\textbf{a},\textbf{a'}, \ref{fig:supp_ISIs}). In fact, the ISI distributions are mainly determined by the mean firing rate.
This result was further supported by coefficients of variation  close to unity, as expected for exponential ISI distributions and Poisson firing (Fig. \ref{fig:supp_ISIs}).

\begin{figure}
 \centering
\includegraphics[width=\textwidth]{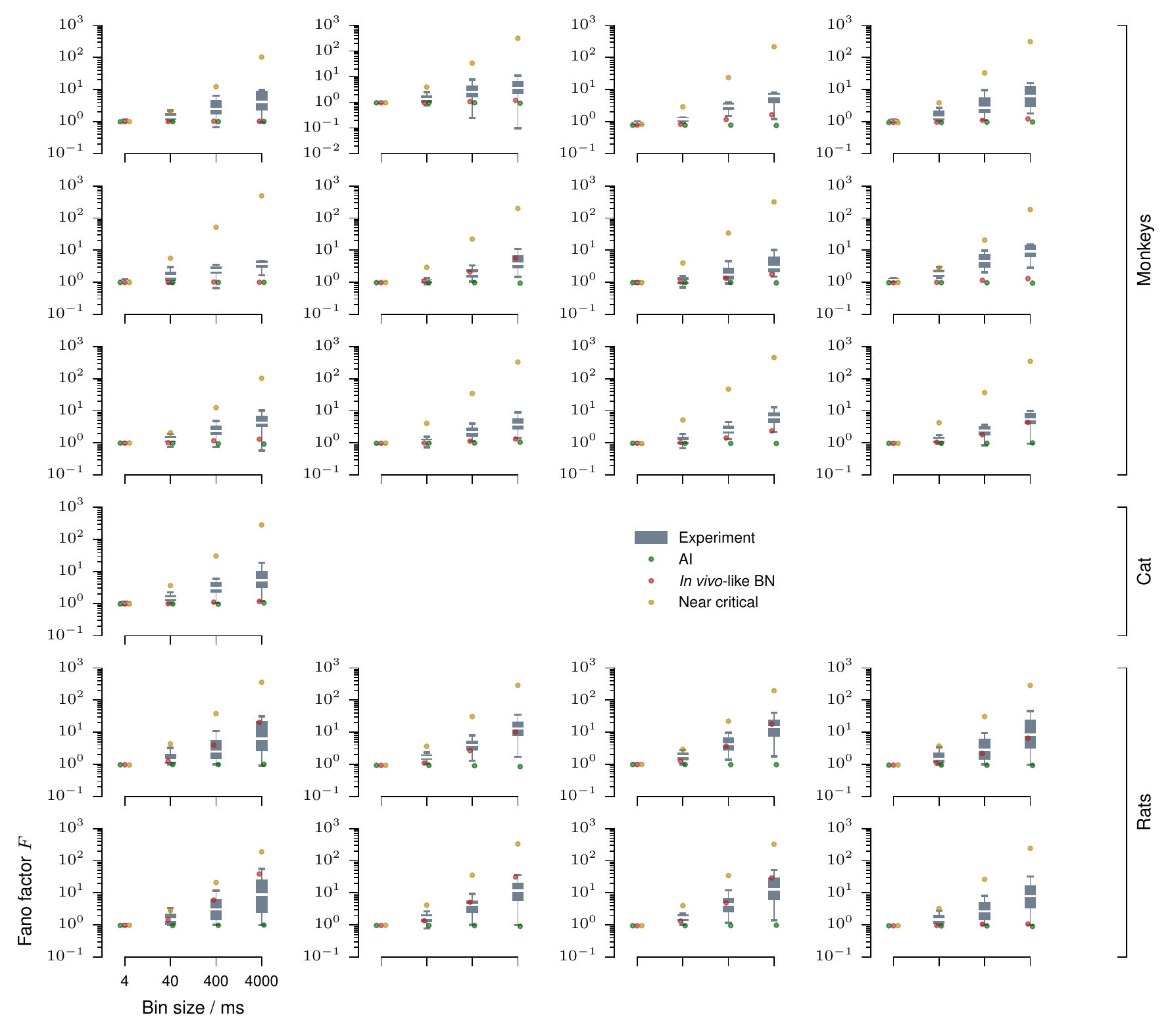}
\caption{\textbf{Fano factors for individual recording sessions.} Fano factors are shown for individual single or multi units of every recording (gray boxplots, median / 25\% -- 75\%, 2.5\% -- 97.5\%), as well as for the matched models, either AI (green), \textit{in vivo}-like (red), or near critical (yellow).}
\label{fig:supp_exp_fanos}
\end{figure}

Second, for both the AI and reverberating regime, the Fano factor $F$ for single unit activity was close to unity, a hallmark feature of irregular spiking \change{\autocite{Tolhurst1981,Vogels1989,Softky1993,Gur1997,Steveninck1997,Kara2000,Carandini2004}} (Fig. \ref{fig:predictions}\textbf{g}, analytical result:  Eq. \eqref{eq:supp_hypergeom_subsampled_fano}).
Hence it cannot serve to distinguish between these different dynamical states.
When evaluating more units, or increasing the bin size to \SI{4}{s}, the differences became more pronounced, but for experiments, the median Fano factor of single unit activity did not exceed $F = 10$ in any of the experiments, even in those with the longest reverberation (Figs. \ref{fig:animals}\textbf{b},\textbf{b'}, \ref{fig:supp_exp_fanos}).
In contrast, for the full network the Fano factor rose to $F \approx 10^4$ for the \invivo-like branching model and diverged when approaching criticality (Fig. \ref{fig:predictions}\textbf{g}, analytical result: Eq. \eqref{eq:fano_factor}).

\begin{figure}
 \centering
\includegraphics[width=\textwidth]{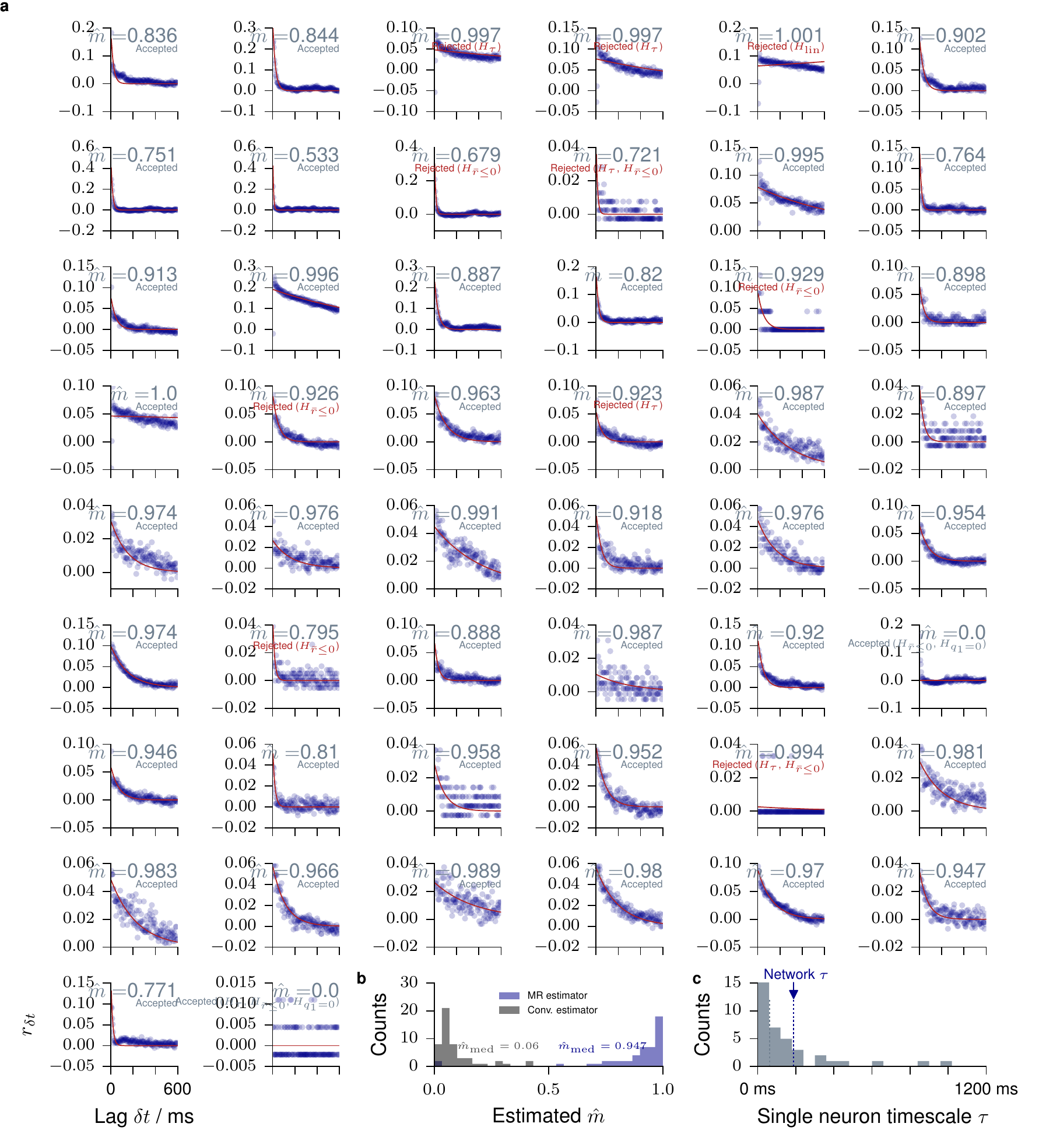}
\caption{\textbf{MR estimation from single neuron activity (cat).} Modified from \cite{Wilting2018}.
MR estimation is used to estimate $\hat{m}$ from the activity $a_t$ of a single units in cat visual cortex. \textbf{a.} Each panel shows MR estimation for one of the 50 recorded units. Autocorrelations decay rapidly in some units, but long-term correlations are present in the activity of most units. The consistency checks are detailed in \cite{Wilting2018}. \textbf{b.} Histogram of the single unit branching ratios $\hat{m}$, inferred with the conventional estimator and using MR estimation. The difference between these estimates demonstrates the subsampling bias of the conventional estimator, and how it is overcome by MR estimation.
\textbf{c}. Histogram of single unit timescales with their median (gray dotted line) and the timescale of the dynamics of the whole network (blue dotted line).}
\label{fig:supp_cat_single_electrodes}
\end{figure}

Third, conventional regression estimators \autocite{Heyde1972,Wei1991} are biased towards inferring irregular activity, as shown before.
Here, conventional estimation yielded a median of $\mh = 0.057$ for single neuron activity in cat visual cortex, in contrast to $\mh = 0.954$ returned by MR estimation  (Fig. \ref{fig:supp_cat_single_electrodes}).

Fourth, for the autocorrelation function of an experimental recording (Fig. \ref{fig:autocorr}\textbf{b}) the rapid decay of $r_{\delta t}$ prevails, and hence single neuron activity appears uncorrelated in time.

\subsection*{Cross-validation of model predictions}
We compared the experimental results to an \textit{in vivo}-like model, which was matched to each experiment only in the average firing rate, and in the inferred branching ratio $\mh$.
Remarkably,
this \textit{in vivo}-like branching model could predict statistical properties not only of single neurons (ISI and Fano factor, see above), but also pairwise and population properties, as detailed below.
This prediction capability further underlines the usefulness of this simple model to approximate the \change{default} state of cortical dynamics.

\begin{figure}
 \centering
\includegraphics[width=\textwidth]{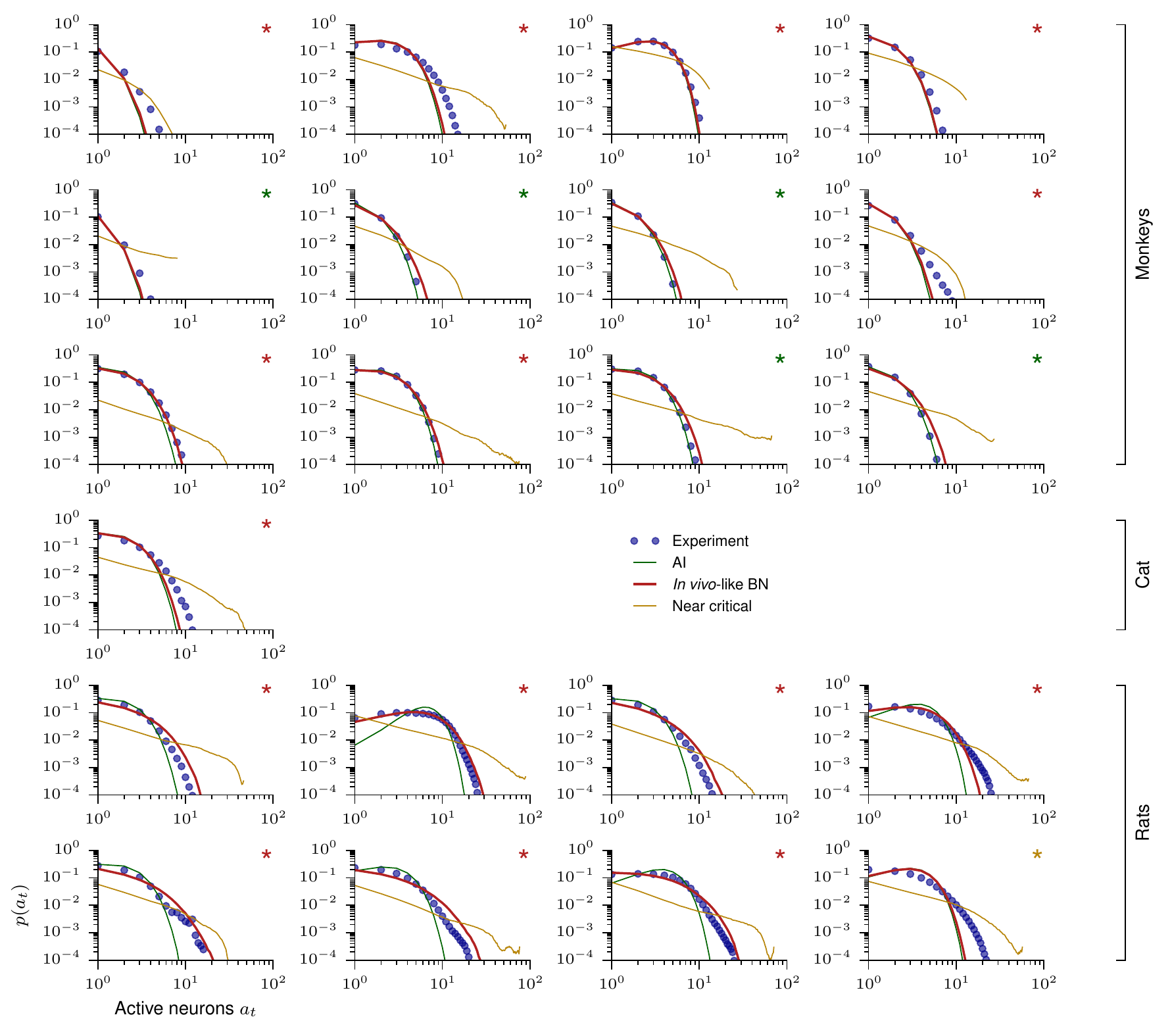}
\caption{\textbf{Activity distributions (4 ms bin size).} Activity distributions are shown for every recording for a bin size of 4 ms (blue). Activity distributions for the matched models, either AI (green), \textit{in vivo}-like (red), or near critical (yellow) are also shown. The color of the asterisk indicates which of the three models yielded the highest likelihood for the data following \cite{Clauset2009}.}
\label{fig:supp_dists_4}
\end{figure}

\begin{figure}
 \centering
\includegraphics[width=\textwidth]{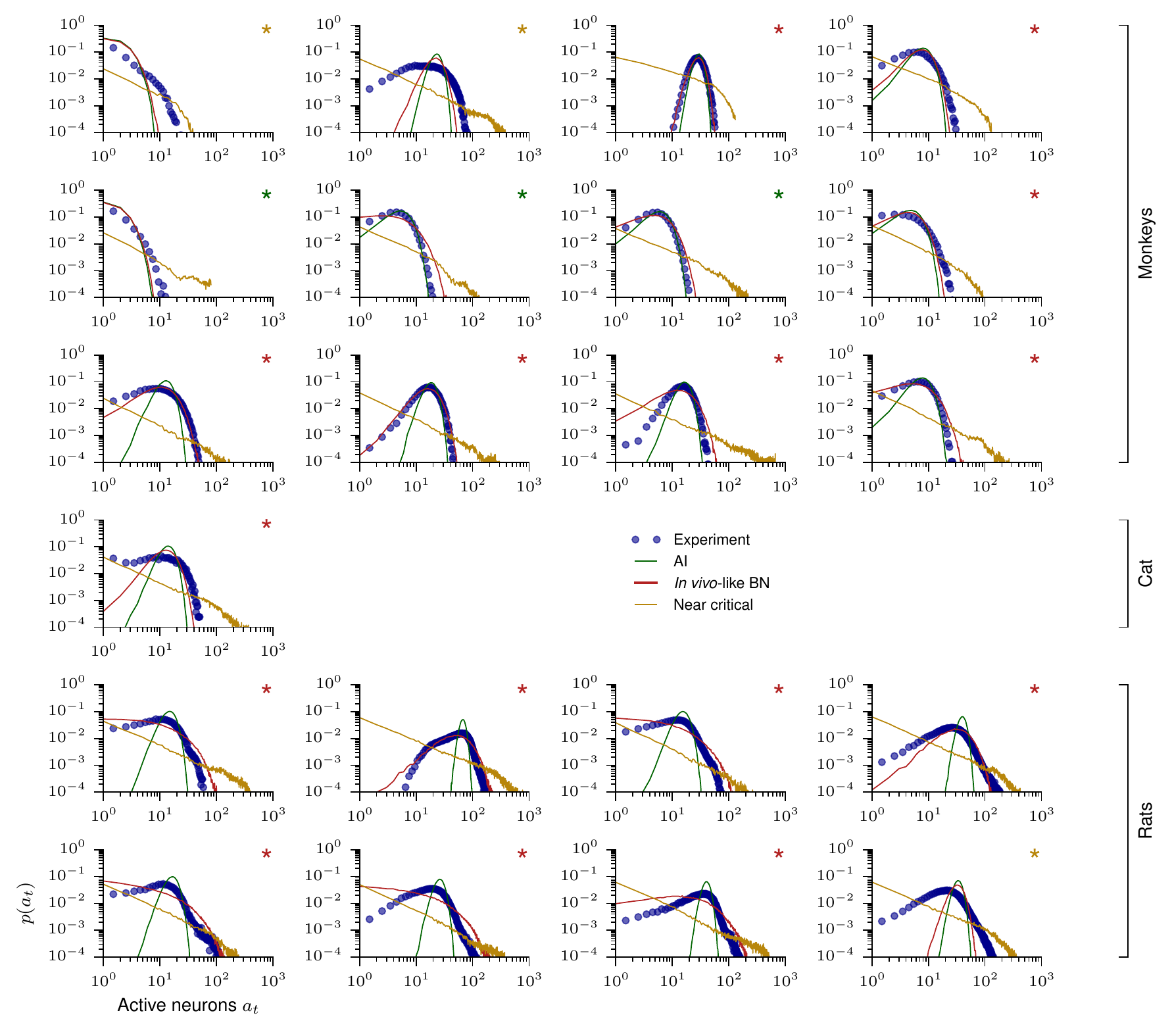}
\caption{\textbf{Activity distributions (40 ms bin size).} Activity distributions are shown for every recording, for a bin size of 40 ms (blue). Activity distributions for the matched models, either AI (green), \textit{in vivo}-like (red), or near critical (yellow) are also shown. The color of the asterisk indicates which of the three models yielded the highest likelihood for the data following \cite{Clauset2009}.}
\label{fig:supp_dists_40}
\end{figure}

First, the model predicted the activity distributions, $p(a_t)$, better than AI or critical models for the majority of experiments (15 out 21, Figs. \ref{fig:animals}\textbf{c},\textbf{d},\textbf{c'},\textbf{d'}, \ref{fig:supp_dists_4}, \ref{fig:supp_dists_40}), both for the exemplary bin sizes of \SI{4}{ms} and \SI{40}{ms}.
Hence, the branching models, which were only matched in their respective first moment of the activity distributions (through the rate) and first moment of the spreading behavior (through $m$), in fact approximated all higher moments of the activity distributions $p(a_t)$.

\begin{figure}
 \centering
\includegraphics[width=\textwidth]{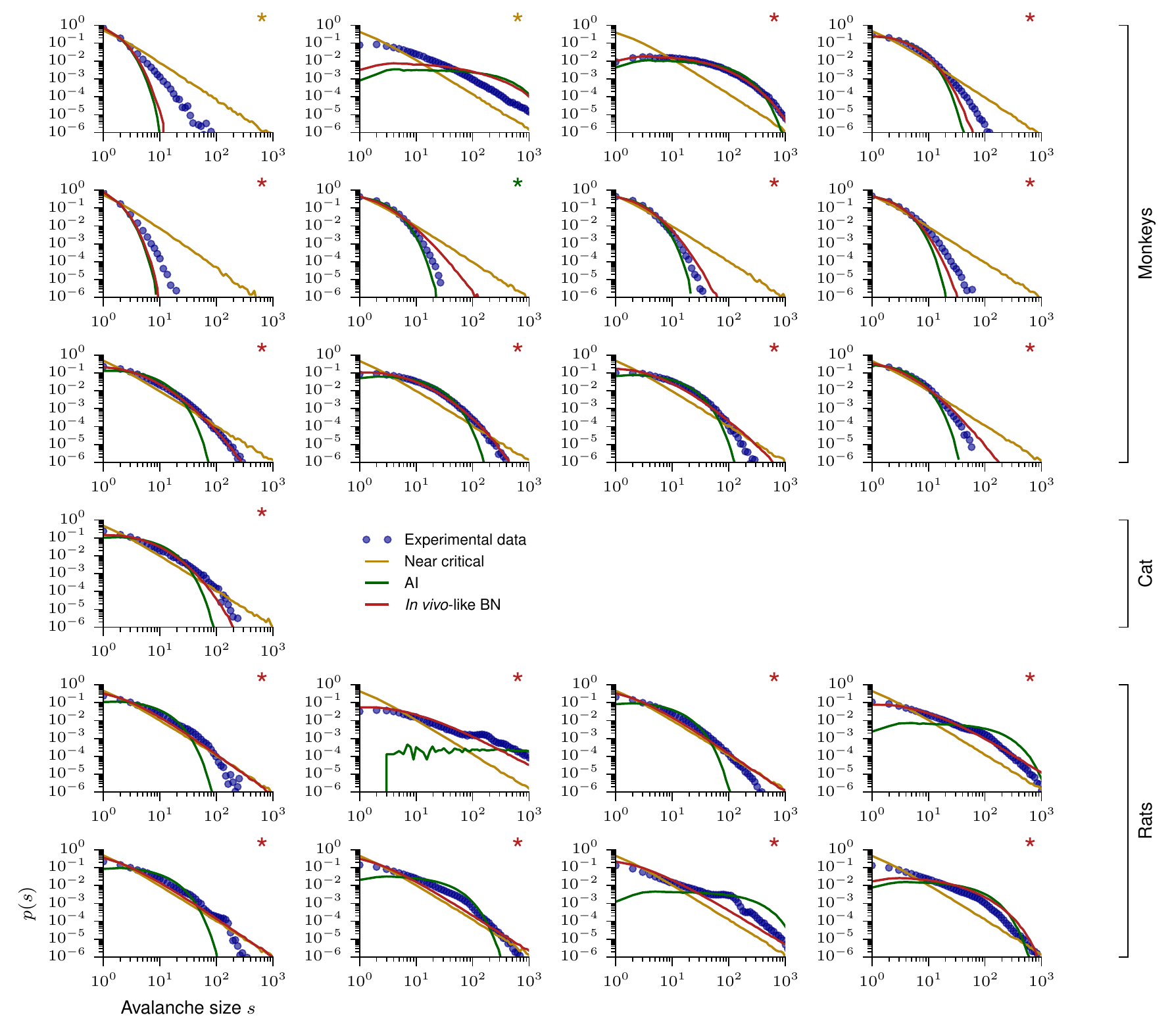}
\caption{\textbf{Avalanche size distribution for individual recording sessions.} Avalanche size distributions are shown for every recording (blue) and for matched models, either AI (green), \textit{in vivo}-like (red), or near critical (yellow).  The color of the asterisk indicates which of the three models yielded the highest likelihood for the data following \cite{Clauset2009}.}
\label{fig:supp_avalanches}
\end{figure}

\begin{figure}
 \centering
\includegraphics[width=\textwidth]{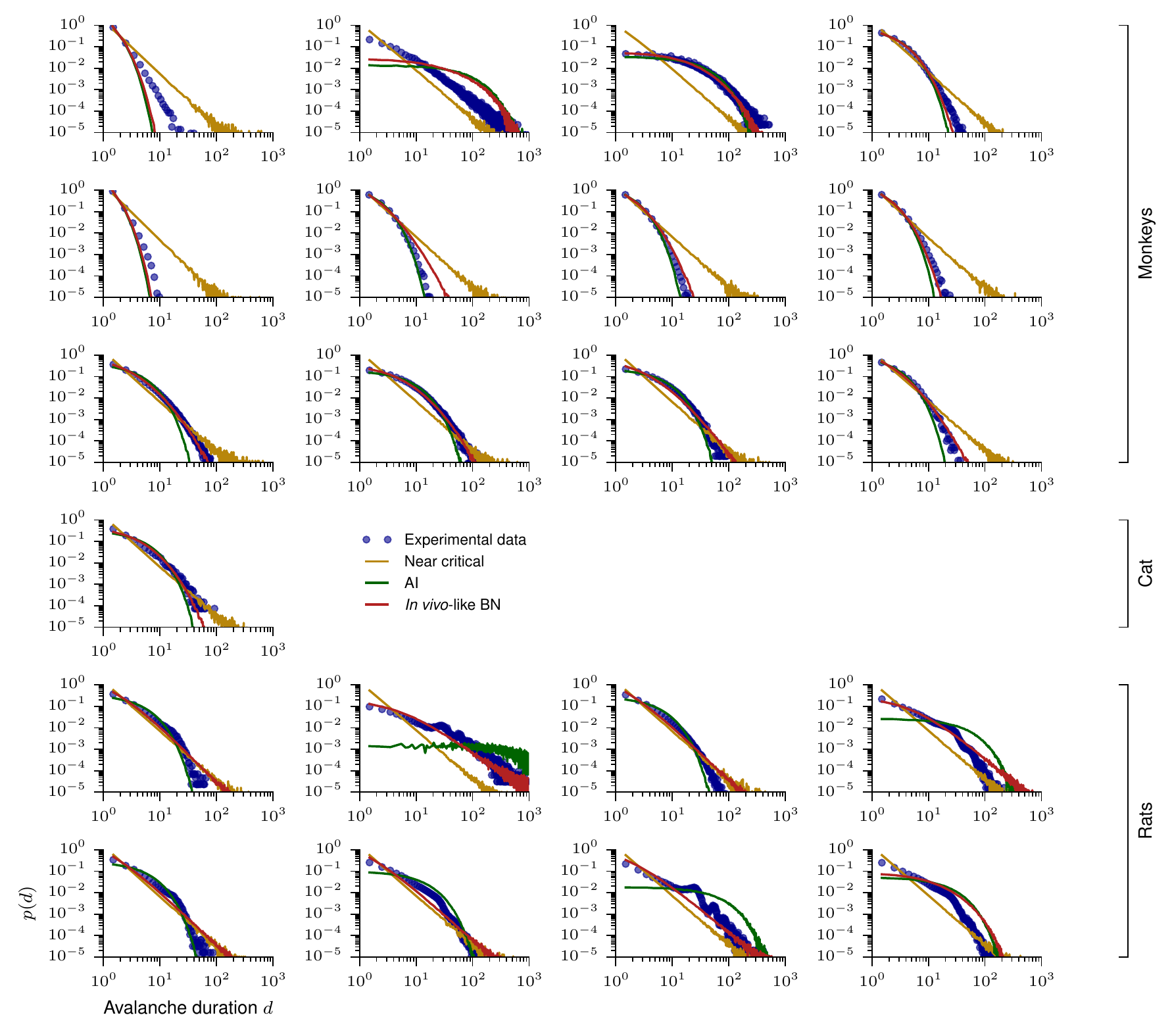}
\caption{\textbf{Avalanche duration distribution for individual recording sessions.} Avalanche duration distributions are shown for every recording (blue) and for matched models, either AI (green), \textit{in vivo}-like (red), or near critical (yellow).}
\label{fig:supp_avalanche_durations}
\end{figure}

Likewise, the model predicted the distributions of neural avalanches, i.e. spatio-temporal clusters of activity (Figs. \ref{fig:animals}\textbf{e},\textbf{f},\textbf{e'}, \textbf{f'}, \ref{fig:supp_avalanches}, \ref{fig:supp_avalanche_durations}).
Characterizing these distributions is a classic approach to assess criticality in neuroscience \autocite{Beggs2003, Priesemann2014}, because avalanche size and duration distributions, $p(s)$ and $p(d)$, respectively, follow power laws in critical systems.
In contrast, for AI activity, they are approximately exponential \autocite{Priesemann2018}.
The matched branching models predicted neither exponential nor power law distributions for the avalanches, but very well matched the experimentally obtained distributions (compare red and blue in Figs. \ref{fig:animals}\textbf{e},\textbf{f},\textbf{e'}, \textbf{f'}, \ref{fig:supp_avalanches}, \ref{fig:supp_avalanche_durations})).
Indeed, model likelihood \autocite{Clauset2009} favored the \textit{in vivo}-like branching model over Poisson and critical models for the majority experiments (18 out of 21, Fig. \ref{fig:supp_avalanches}).
Our results here are consistent with those of spiking activity in awake animals, which typically do not display power laws \autocite{Priesemann2014, Ribeiro2010, Bedard2006}.
In contrast, most evidence for criticality \change{\textit{in vivo}}, in particular the characteristic power-law distributions, has been obtained from \textit{coarse} measures of neural activity (LFP, EEG, BOLD; see \cite{Priesemann2014} and references therein).

\begin{figure}
 \centering
\includegraphics[width=\textwidth]{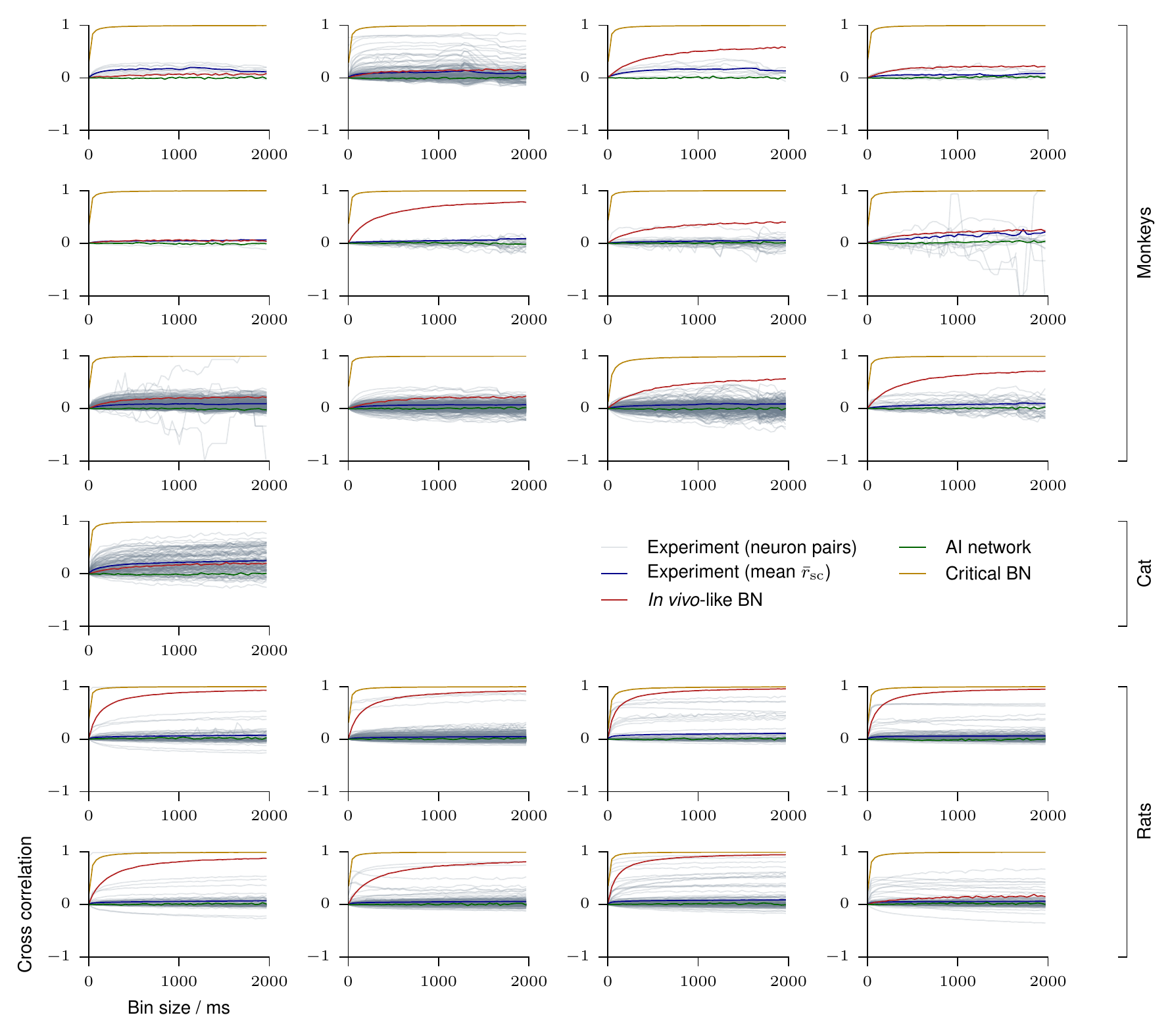}
\caption{\textbf{Cross correlations for individual recording sessions.} Spike count cross correlations ($r_\mathrm{sc}$) are shown for every neuron pair (gray) and the ensemble average (blue) of each recording, for bin sizes from 1 ms to 2s. Cross correlations are also shown for the matched models, either AI (green), \textit{in vivo}-like (red), or near critical (yellow).}
\label{fig:supp_corrs}
\end{figure}

Last, the model predicted the pairwise spike count cross correlation ${r}_\mathrm{sc}$.
In experiments, ${r}_\mathrm{sc}$ is typically between 0.01 and 0.25, depending on brain area, task, and most importantly, the analysis timescale (bin size) \autocite{Cohen2011}.
For the cat recording the model even correctly predicted the bin size dependence of ${r}_\mathrm{sc}$ from $\bar{r}_\mathrm{sc} \approx 0.004$ at a bin size of \SI{4}{ms} (analytical result: Eq. \eqref{eq:supp_spike_count_correlation})  to $\bar{r}_\mathrm{sc} \approx 0.3$ at a bin size of \SI{2}{s} (Fig. \ref{fig:animals}\textbf{g}).
Comparable results were also obtained for some monkey experiments.
In contrast, correlations in most monkey and rat recordings were smaller than predicted (Figs. \ref{fig:animals}\textbf{g'}, \ref{fig:supp_corrs}).
It is very surprising that the model correctly predicted the cross-correlation even in some experiments, as $m$ was inferred only from the \textit{temporal} structure of the spiking activity alone, whereas ${r}_\mathrm{sc}$ characterizes spatial dependencies.

Overall, by only estimating the effective synaptic strength $m$ from the \textit{in vivo} recordings, higher-order properties like avalanche size distributions, activity distributions and in some cases spike count cross correlations could be closely matched using the generic branching model.

\subsection*{The dynamical state determines responses to small stimuli}
\begin{figure}
 \centering
\includegraphics[width=\textwidth]{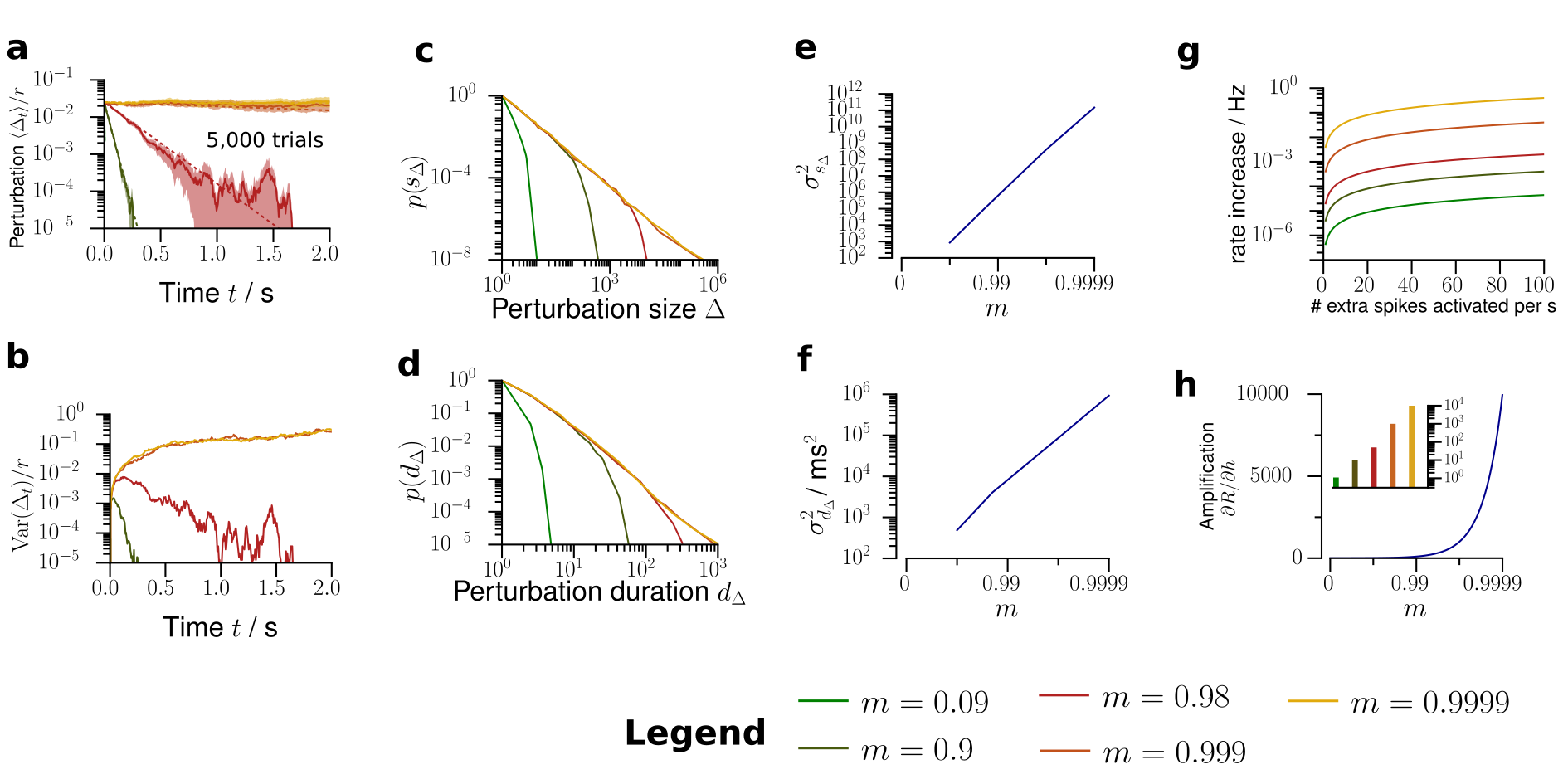}
\caption{\textbf{Further predictions about network activity.} 
\textbf{a}. The model predicts that the perturbation decays exponentially with decay time $\tau = - \Delta t / \log m$.
\textbf{b} The variance across trials of the perturbed firing rate has a maximum, whose position depends on $m$.
\textbf{c}. Depending on $m$, the model predicts the distributions for the total number of extra spikes $s_\Delta$ generated by the network following a single extra spike.
\textbf{d}. Likewise, the model predicts distributions of the duration $d$ of these perturbations.
\textbf{e}. Variance of the total perturbation size as a function of $m$.
\textbf{f}. Variance of the total perturbation duration as a function of $m$.
\textbf{g}. Increase of the network firing rate as a function of the rate of extra neuron activations for different $m$.
\textbf{h}. Amplification (susceptibility) $\mathrm{d} r / \mathrm{d} h$ of the network as a function of the branching ratio $m$.
}
\label{fig:supp_predictions}
\end{figure}

After validating the model using a set of statistical properties that are well accessible experimentally, we now turn to making predictions about yet unknown properties, namely network responses to small stimuli.
In the line of \cite{London2010}, assume that on a background of spiking activity one single extra spike is triggered.
This spike may in turn trigger new spikes, leading to a cascade of additional spikes $\Delta_t$ propagating through the network.
A dynamical state with branching ratio $m$ implies that \textit{on average}, this perturbation decays with time constant $\tau = - \Delta t / \log m$.
Similar to the approach in \cite{London2010}, the evolution of the mean firing rate, averaged over a reasonable number of trials (here: 500) unveils the nature of the underlying spike propagation: depending on $m$, the rate excursions will last longer, the higher $m$ (Figs. \ref{fig:predictions}\textbf{a},\textbf{b},\textbf{c}, \ref{fig:supp_predictions}\textbf{a}).
The perturbations are not deterministic, but show trial-to-trial variability which also increases with $m$ (\ref{fig:supp_predictions}\textbf{b}).

Unless $m>1$, the theory of branching models ensures that perturbations will die out eventually after a duration $d_\Delta$, having accumulated a total of $\change{s_\Delta} = \sum_{t=1}^d \Delta_t$ extra spikes in total.
This perturbation size $\change{s_\Delta}$ and duration $\change{d_\Delta}$ follow specific distributions, \autocite{Harris1963} which are determined by $m$: they are power law distributed in the critical state ($m=1$), with a cutoff for any $m<1$ (Figs. \ref{fig:predictions}\textbf{f},  \ref{fig:supp_predictions}\textbf{c},\textbf{d}).
These distributions imply a characteristic mean perturbation size $\E{s_\Delta}$ (Fig. \ref{fig:predictions}\textbf{d}), which diverges at the critical point.
The variability of the perturbation sizes is also determined by $m$ and also diverges at the critical point (inset of Fig. \ref{fig:predictions}\textbf{d}, Fig. \ref{fig:supp_predictions}\textbf{e}).

Taken together, these results imply that the closer a neuronal network is to criticality, the more sensitive it is to external perturbations, and the better it can amplify small stimuli.
At the same time, these networks also show larger trial-to-trial variability.
For typical cortical networks, we found that the response to one single extra spike will on average comprise between 20 and 1000 additional spikes in total (Figs. \ref{fig:predictions}\textbf{e}).

\subsection*{The dynamical state determines network susceptibility and variability}
Moving beyond single spike perturbations, our model gives precise predictions for the network response to continuous stimuli.
If extra action potentials are triggered at rate $h$ in the network, the network will again amplify these external activations, depending on $m$.
Provided an appropriate stimulation protocol, this rate response could be measured and our prediction tested in experiments (Fig. \ref{fig:supp_predictions}\textbf{g}).
The susceptibility \change{$\partial R / \partial h$} diverges at the critical transition and is unique to a specific branching ratio $m$.
We predict that typical cortical networks will amplify a small, but continuous input rate by about a factor fifty (Fig. \ref{fig:supp_predictions}\textbf{h}, red).

While the input and susceptibility determine the network's mean activity, the network still shows strong rate fluctuations around this mean value.
The magnitude of these fluctuations in relation to the mean can be quantified by the network Fano factor $F = \Var{A_t} \, / \, \E{A_t}$ (Fig. \ref{fig:predictions}\textbf{g}).
This quantity cannot be directly inferred from experimental recordings, because the Fano factor of subsampled populations severely underestimates the network Fano factor, as shown before.
We here used our \textit{in vivo}-like model to obtain estimates of the network Fano factor: for a bin size of $\Delta t = \SI{4}{ms}$ it is about $F\approx 40$ and rises to $F\approx 4000$ for bin sizes of several seconds, highlighting that network fluctuations probably are much stronger than one would naively assume from experimental, subsampled spiking activity.

\subsection*{Distinguishing afferent and recurrent activation}
\begin{figure*}[h!]
\includegraphics[width=\textwidth]{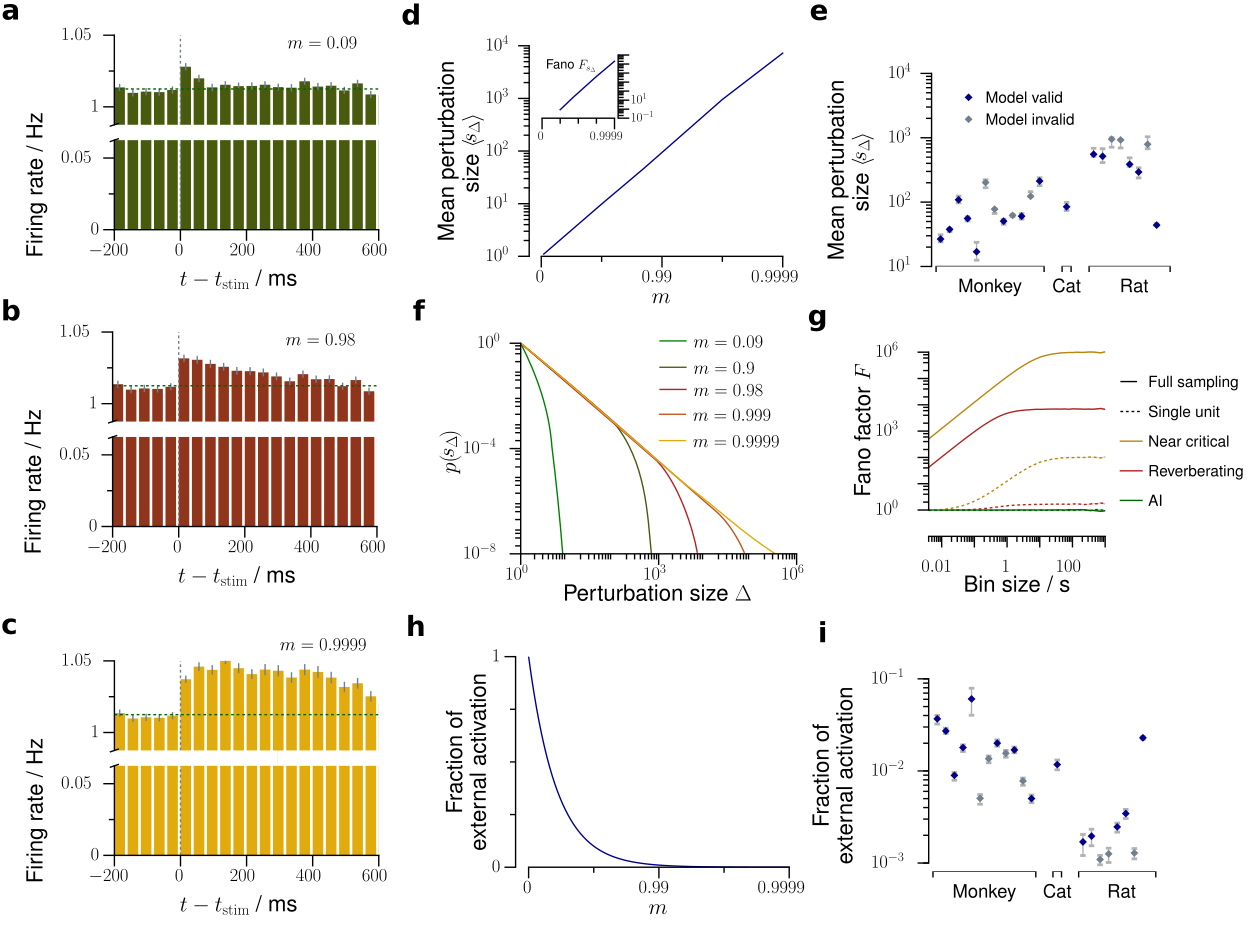}
\caption{\textbf{Predictions about network dynamics and propagation of perturbations.}
Using our \textit{in-vivo}-like, reverberating model, we can predict several network properties, which are yet very complicated or impossible to obtain experimentally.
\textbf{a -- c.} In response to one single extra spike, a perturbation propagates in the network depending on the branching ratio $m$, and can be observed as a small increase of the average firing rate of the sampled neurons, here simulated for 500 trials (as in \cite{London2010}). This increase of firing rate decays exponentially, with the decay time $\tau$ being determined by $m$. The perturbation \textbf{a} is rapidly quenched in the asynchronous-irregular state, \textbf{b} decays slowly over hundreds of milliseconds in the reverberating state, or \textbf{c} persists almost infinitely in the critical state.
    \textbf{d}. The average perturbation size $\E{s_\Delta}$ and Fano factor $F_{s_\Delta}$ (inset) increase strongly with $m$.
\textbf{e}. Average total perturbation sizes predicted for each spike recording of mammalian cortex (errorbars: 5\% -- 95\% confidence intervals).
\textbf{f}. Distribution $p(s_\Delta)$ of total perturbation sizes $s_\Delta$. The asynchronous-irregular models show approximately Poisson distributed, near critical models power-law distributed perturbation sizes.
\textbf{g}. Bin size dependent Fano factors of the activity, here exemplarily shown for the asynchronous-irregular ($m=0$, green), representative reverberating ($m=0.98$, red), and near critical ($m=0.9999$, yellow) models. While the directly measurable Fano factor of single neurons (dotted lines) underestimates the Fano factor of the whole network, the model allows to predict the Fano factor of the whole network (solid lines).
\textbf{h}. The fraction of the externally generated spikes compared to all spikes in the network strongly decreases with larger $m$.
\textbf{i}. Fraction of the externally generated spikes predicted for each spike recording of mammalian cortex (errorbars as in \textbf{e}).
}
\label{fig:predictions}
\end{figure*}

Last, our model gives an easily accessible approach to solving the following question: given a spiking neuronal network, which fraction of the activity $\E{A}$ is generated by recurrent activation from within the network, and which fraction can be attributed to external, afferent excitation $h$?
The branching model readily provides an answer: the fraction of external activation
is $h / \E{A} = 1 - m$ (Fig. \ref{fig:predictions}\textbf{h}).
In this framework, AI-like networks are completely driven by external input currents or noise, whereas reverberating networks generate a substantial fraction of their activity intrinsically.
For the experiments investigated in this study, we inferred that between 0.1\% and 7\% of the activity are externally generated (median 2\%, Fig. \ref{fig:predictions}\textbf{i}). 

While our model is quite simplistic given the complexity of neuronal network activity, keep in mind that ``all models are wrong, but some are useful'' \autocite{Box1979}.
Here, the model has proven to provide a good first order approximation to a number of statistical properties of spiking activity and propagation in cortex. 
\change{
Hence, it promises insight into cortical function because (i) it relies on simply assessing spontaneous cortical activity, (ii) it does not require manipulation of cortex, (iii) it enables reasonable predictions about sensitivity, amplification, and internal and external activation, (iv) this analysis is possible in an area specific, task- and state-dependent manner as only short recordings are required for consistent results.
}

\section*{Discussion}

\subsection*{Our results resolve contradictions between AI and critical states}

Our results for spiking activity \textit{in vivo} suggest that network dynamics show AI-like statistics, because under subsampling the observed correlations are underestimated.
In contrast, typical experiments that assessed criticality potentially overestimated correlations by sampling from overlapping populations (LFP, EEG) and thereby hampered a fine distinction between critical and subcritical states \autocite[in prep]{Neto2017}.
By employing for the first time a consistent, quantitative estimation, we provided evidence that \textit{in vivo} spiking population dynamics reflects a reverberating regime, i.e. it operates in a narrow regime around $m=0.98$.
This result is supported by the findings by \cite{Dahmen2016}: based on distributions of covariances, they inferred that cortical networks operate in a regime below criticality.
Given the generality of our results across different species, brain areas, and cognitive states, our results suggest self-organization to this reverberating regime as a general organization principle for cortical network dynamics.

\change{
\subsection*{The reverberating regime combines features of AI and critical state}
At first sight, $\sh=0.98$ of the reverberating regime may suggest that the collective spiking dynamics is very close to critical.
Indeed, physiologically a $\Delta m \approx $ 1.6\% difference to criticality ($m=1$) is small in terms of the effective synaptic strength. 
However, this apparently small difference in single unit properties has a large impact on the collective \emph{dynamical} fingerprint and makes AI, reverberating, and critical states clearly distinct:
For example, consider the sensitivity to a small input, i.e. the susceptibility  \change{$\chi = \partial R \, / \, \partial h = \frac{1}{1 - m}$}.
The susceptibility diverges at criticality, making critical networks overly sensitive to input.
In contrast, states with $m \approx 0.98$ assure sensitivity without instability.
Because this has a strong impact on network dynamics and putative network function, finely distinguishing between dynamical states is both important and feasible even if the corresponding differences in effective synaptic strength ($m$) appear small.
}

\change{
We cannot ultimately rule out that cortical networks self-organize as close as possible towards criticality, the platonic ideal being impossible to achieve for example due to finite-size, external input, and refractory periods.
Therefore, the reverberating regime might conform with quasi-criticality \autocite{Williams-Garcia2014} or neutral theory \autocite{Martinello2017}.
However, we deem this unlikely for two reasons.
First, in simulations of finite-size networks with external input, we could easily distinguish the reverberating regime from states with $m=0.9999$ \autocite{Wilting2018}, which are more than one order of magnitude closer to criticality than any experiment we analyzed.
Second, operating in a reverberating regime, which is between AI and critical, may combine the computational advantages of both states \autocite{Wilting2018a}:
 the reverberating regime enables rapid changes of computational properties by small parameter changes, keeps a sufficient safety-margin from instability to make seizures sufficiently unlikely \autocite{Priesemann2014}, balances competing requirements (e.g. sensitivity and specificity \autocite{Gollo2017}), and may carry short term memory and allow to integrate information over limited, tunable timescales \autocite{Wang2002,Boedecker2012}.
}
\change{
For these reasons, we consider the reverberating regime to be the explicit target state of self-organization.
This is in contrast to the view of ``as close to critical as possible'', which still holds criticality as the ideal target.
}

\subsection*{More complex network models}
Cortical dynamics is clearly more complicated than a simple branching model.
For example, heterogeneity of single-neuron morphology and dynamics, and non-trivial network topology likely impact population dynamics.
However, we showed that \textit{statistics} of cortical network activity are well approximated by a branching model.
Therefore, we interpret branching models as a \textit{statistical} approximation of spike propagation, which can capture a fair extent of the complexity of cortical dynamics.
By using branching models, we draw on the powerful advantage of analytical tractability, which allowed for basic insight into dynamics and stability of cortical networks.

\begin{figure}
 \centering
\includegraphics[width=\textwidth]{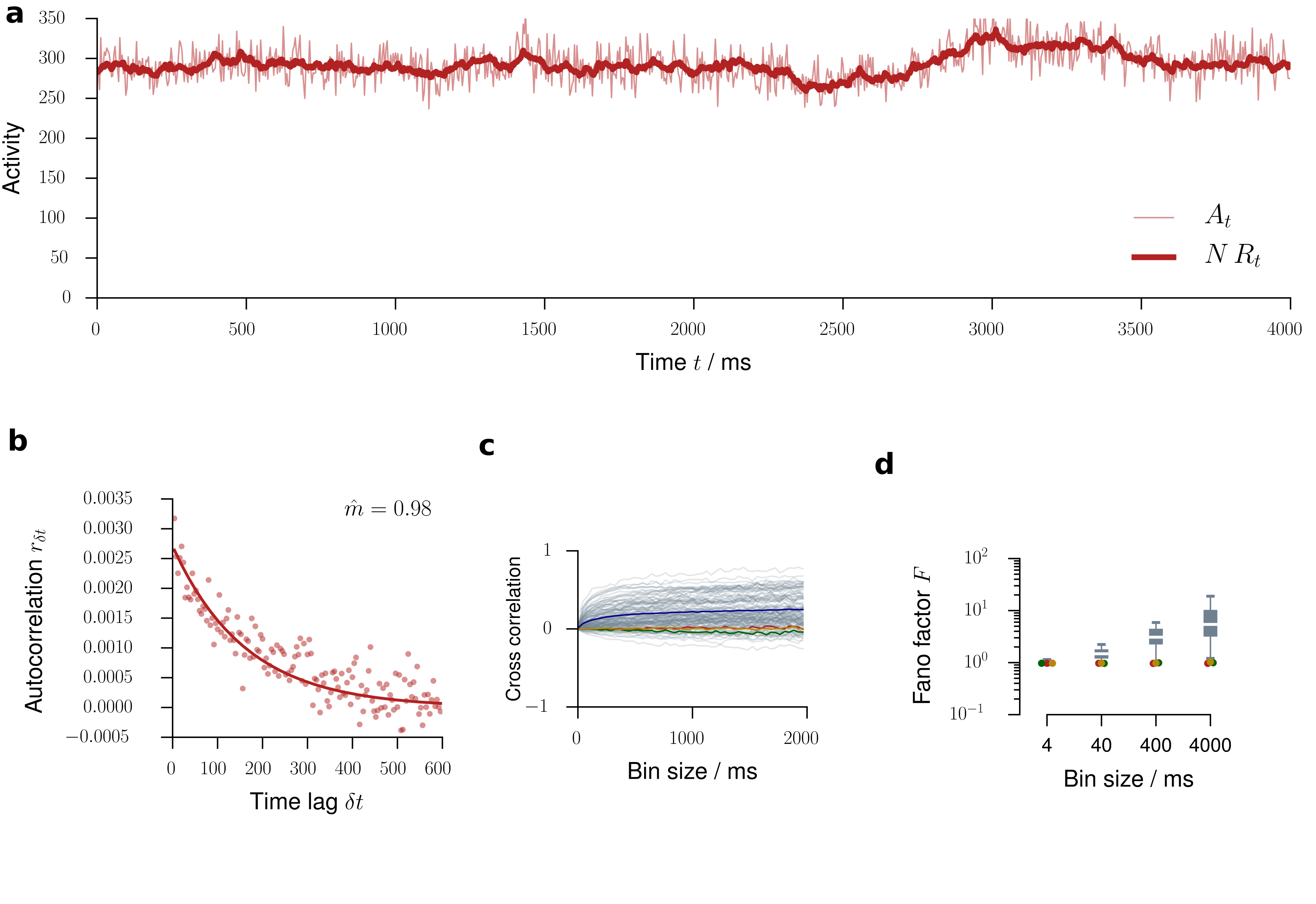}
\caption{\textbf{Doubly stochastic model.}
    Instead of a branching model, we here matched a doubly stochastic process to the data. The rates evolved according to $R_{t+1} = m \, R_t + h_t$ where $h_t$ is drawn from a Poisson distribution. The actual activity is then drawn from a Poisson distribution according to $A_t\sim \mathrm{Poi}(N \, R_t)$. Here, results for the experiment in cat visual cortex are shown.
    \textbf{a} Time evolution of $R_t$ and $A_t$. As the activity is not fed back into the evolution of $R_t$, the second step effectively adds measurement noise to the underlying process.
    \textbf{b} The subsampled activity (50 out of 10,000, as in the branching models) shows the expected autocorrelation function.
    \textbf{c} Any of the doubly stochastic processes underestimated the spike count cross correlations.
    \textbf{d} Any of the doubly stochastic processes underestimated the single unit Fano factors.
    \label{fig:supp_doubly_stochastic}
    }
\end{figure}

\change{
    In contrast to the branching model, doubly stochastic processes (i.e. spikes drawn from an inhomogeneous Poisson distribution) failed to reproduce many statistical features (Fig. \ref{fig:supp_doubly_stochastic}).
    We conjecture that the key difference is that doubly stochastic processes propagate the underlying firing rate instead of the actual spike count.
    Thus, propagation of the actual number of spikes (as e.g. in branching or Hawkes processes \autocite{Kossio2018}), not some underlying firing rate, seems to be integral to capture the statistics of cortical spiking dynamics.
}

\change{
Our statistical model stands in contrast to \textit{generative} models, which generate spiking dynamics by physiologically inspired mechanisms.
One particularly prominent example are networks with balanced excitation and inhibition \autocite{Vreeswijk1996a,VanVreeswijk1997,Brunel2000a}, which became a standard model of neuronal networks \autocite{Hansel2012}.
A balance of excitation and inhibition is supported by experimental evidence \autocite{Okun2008}.
Our statistical model reproduces statistical properties of such networks if one assumes that the excitatory and inhibitory contributions can be described by an effective excitation.
In turn, the results obtained from the well-understood estimator can guide the refinement of generative models.
For example, we suggest that network models need to be extended beyond the asynchronous-irregular state \autocite{Brunel2000a} to incorporate the network reverberations observed \textit{in vivo}.
Possible candidate mechanisms are increased coupling strength or inhomogeneous connectivity.
Both have already been shown to induce rate fluctuations with timescales of several hundred milliseconds \autocite{Litwin-kumar2012,Ostojic2014,Kadmon2015}.
}

\change{
Because of the assumption of uncorrelated, Poisson-like network firing, models that study single neurons typically assume that synaptic currents are normally distributed. 
Our results suggest that one should rather use input with reverberating properties with timescales of a few hundred milliseconds to reflect input from cortical neurons \textit{in vivo}.
This could potentially change our understanding of single neuron dynamics, for example of their input-output properties.
}



\subsection*{Deducing network properties from the tractable model}

Using our analytically tractable model, we could predict and validate network properties, such as avalanche size and duration, interspike interval, or activity distributions.
Given the experimental agreement with these predictions, we deduced further properties, which are impossible or difficult to assess experimentally and gave insight into more complex questions about network responses: how do perturbations propagate within the network, and how susceptible is the network to external stimulation?

One particular question we could address is the following: which fraction of network activity is attributed to external or recurrent, internal activation?
We inferred that about 98\% of the activity is generated by recurrent excitation, and only about 2\% originates from input or spontaneous threshold crossing.
This result may depend systematically on the brain area and cognitive state investigated:
For layer 4 of primary visual cortex in awake mice, \cite{Reinhold2015} concluded that the fraction of recurrent cortical excitation is about 72\%, and cortical activity dies out with a timescale of about \SI{12}{ms} after thalamic silencing.
Their numbers agree perfectly well with our phenomenological model: a timescale of $\tau=$\SI{12}{ms} implies that the fraction of recurrent cortical excitation is 
$m = e^{-\Delta t / \tau} \approx 72\%$, just as found experimentally.
Under anesthesia, in contrast, they report timescales of several hundred milliseconds, in agreement with our results.
These differences show that the fraction of external activation may strongly depend on cortical area, layer, and cognitive state.
The novel estimator can in future contribute to a deeper insight into these differences, because it allows for a straight-forward assessment of afferent versus recurrent activation, simply from evaluating spontaneous spiking activity, without the requirement of thalamic or cortical silencing.


\section*{Acknowledgments}
JW received support from the Gertrud-Reemstma-Stiftung.
VP received financial support from the German Ministry for Education and Research (BMBF) via the Bernstein Center for Computational Neuroscience (BCCN) G\"ottingen under Grant No. 01GQ1005B, and by the German-Israel-Foundation (GIF) under grant number G-2391-421.13.
JW and VP received financial support from the Max Planck Society.

\section*{Competing interests}
The authors declare that the research was conducted in the absence of any commercial or financial relationships that could be construed as a potential conflict of interest.

\begin{scriptsize}

\end{scriptsize}

\beginsupplement


\onecolumn

\setcounter{page}{1}

\section*{Supplementary material for ``Between perfectly critical and fully irregular: a reverberating model captures and predicts cortical spike propagation'' by J. Wilting and V. Priesemann}

\subsection{Experiments}
\label{sec:supp_experiments}
We evaluated spike population dynamics from recordings in rats, cats and monkeys.
 The rat experimental protocols were approved by the Institutional Animal Care and Use Committee of Rutgers University \autocite{Mizuseki2009,Mizuseki2009a}.
 The cat experiments were performed in accordance with guidelines established by the Canadian Council for Animal Care \autocite{Blanche2009}.
The monkey experiments were performed according to the German Law for the Protection of Experimental Animals, and were approved by the Regierungspr\"asidium Darmstadt.
The procedures also conformed to the regulations issued by the NIH and the Society for Neuroscience.
The spike recordings from the rats and the cats were obtained from the NSF-founded CRCNS data sharing website \autocite{Blanche2006,Blanche2009,Mizuseki2009,Mizuseki2009a}.

\paragraph*{Rat experiments.}
In rats the spikes were recorded in CA1 of the right dorsal hippocampus during an open field task.
We used the first two data sets of each recording group (ec013.527, ec013.528, ec014.277, ec014.333, ec015.041, ec015.047, ec016.397, ec016.430).
The data-sets provided sorted spikes from 4 shanks (ec013) or 8 shanks (ec014, ec015, ec016), with 31 (ec013), 64 (ec014, ec015) or 55 (ec016) channels. We used both, spikes of single and multi units, because knowledge about the identity and the precise number of neurons is not required for the MR estimator.
More details on the experimental procedure and the data-sets proper can be found in \cite{Mizuseki2009,Mizuseki2009a}.

\paragraph*{Cat experiments.}
Spikes in cat visual cortex were recorded by Tim Blanche in the laboratory of Nicholas Swindale, University of British Columbia \autocite{Blanche2009}.
We used the data set pvc3, i.e. recordings of 50 sorted single units in area 18 \autocite{Blanche2006}.
We used that part of the experiment in which no stimuli were presented, i.e., the spikes reflected spontaneous activity in the visual cortex of the anesthetized cat. Because of potential non-stationarities at the beginning and end of the recording, we omitted data before \SI{25}{s} and after \SI{320}{s} of recording.
Details on the experimental procedures and the data proper can be found in \cite{Blanche2009,Blanche2006}.

\paragraph*{Monkey experiments.}
The monkey data are the same as in \cite{Pipa2009, Priesemann2014}. In these experiments, spikes were recorded simultaneously from up to 16 single-ended micro-electrodes ($ \diameter  = 80 \, \mu  \mathrm{m}$) or tetrodes ($ \diameter  = 96 \, \mu  \mathrm{m}$) in lateral prefrontal cortex of three trained macaque monkeys (M1: 6 kg \female ; M2: 12 kg \male ; M3: 8 kg \female ).
The electrodes had impedances between 0.2 and $1.2 \, \mathrm{M} \Omega$ at 1 kHz, and were arranged in a square grid with inter electrode distances of either 0.5 or 1.0 mm.
The monkeys performed a visual short term memory task. The task and the experimental procedure is detailed in \cite{Pipa2009}.
We analyzed spike data from 12 experimental sessions comprising almost 12.000 trials (M1: 5 sessions; M2: 4 sessions; M3: 3 sessions).
6 out of 12 sessions were recorded with tetrodes.
Spike sorting on the tetrode data was performed using a Bayesian optimal template matching approach as described in \cite{Franke2010} using the “Spyke Viewer” software \autocite{Propper2013}. 
On the single electrode data, spikes were sorted with a multi-dimensional PCA method (Smart Spike Sorter by Nan-Hui Chen).

\subsection{Analysis}
\label{sec:supp_analysis}

\paragraph*{Temporal binning.}
For each recording, we collapsed the spike times of all recorded neurons into one single train of population spike counts $a_{t}$, where $a_{t}$ denotes how many neurons spiked in the $t^{th}$ time bin $\Delta t$.
If not indicated otherwise, we used $\Delta t = \SI{4}{ms}$, reflecting the propagation time of spikes from one neuron to the next.

\paragraph*{Multistep regression estimation of $\mh$.}
From these time series, we estimated $\hat{m}$ using the MR estimator described in \cite{Wilting2018}. For $k=1,\ldots,k_\mathrm{max}$, we calculated the linear regression slope $r_{k\, \Delta t}$ for the linear statistical dependence of $a_{t+k}$ upon $a_t$. From these slopes, we estimated $\mh$ following the relation $r_{\delta t} = b \cdot \mh^{\delta t / \Delta t}$, where $b$ is an (unknown) parameter that depends on the higher moments of the underlying process and the degree of subsampling. However, for an estimation of $m$ no further knowledge about $b$ is required.

Throughout this study we chose $k_\mathrm{max} = 2500$ (corresponding to \SI{10}{s}) for the rat recordings, $k_\mathrm{max} = 150$ (\SI{600}{ms}) for the cat recording, and $k_\mathrm{max} = 500$ (\SI{2000}{ms}) for the monkey recordings, assuring that $k_\mathrm{max}$ was always in the order of multiple intrinsic network timescales (i.e., autocorrelation times). 

\begin{figure}
 \centering
\includegraphics[width=\textwidth]{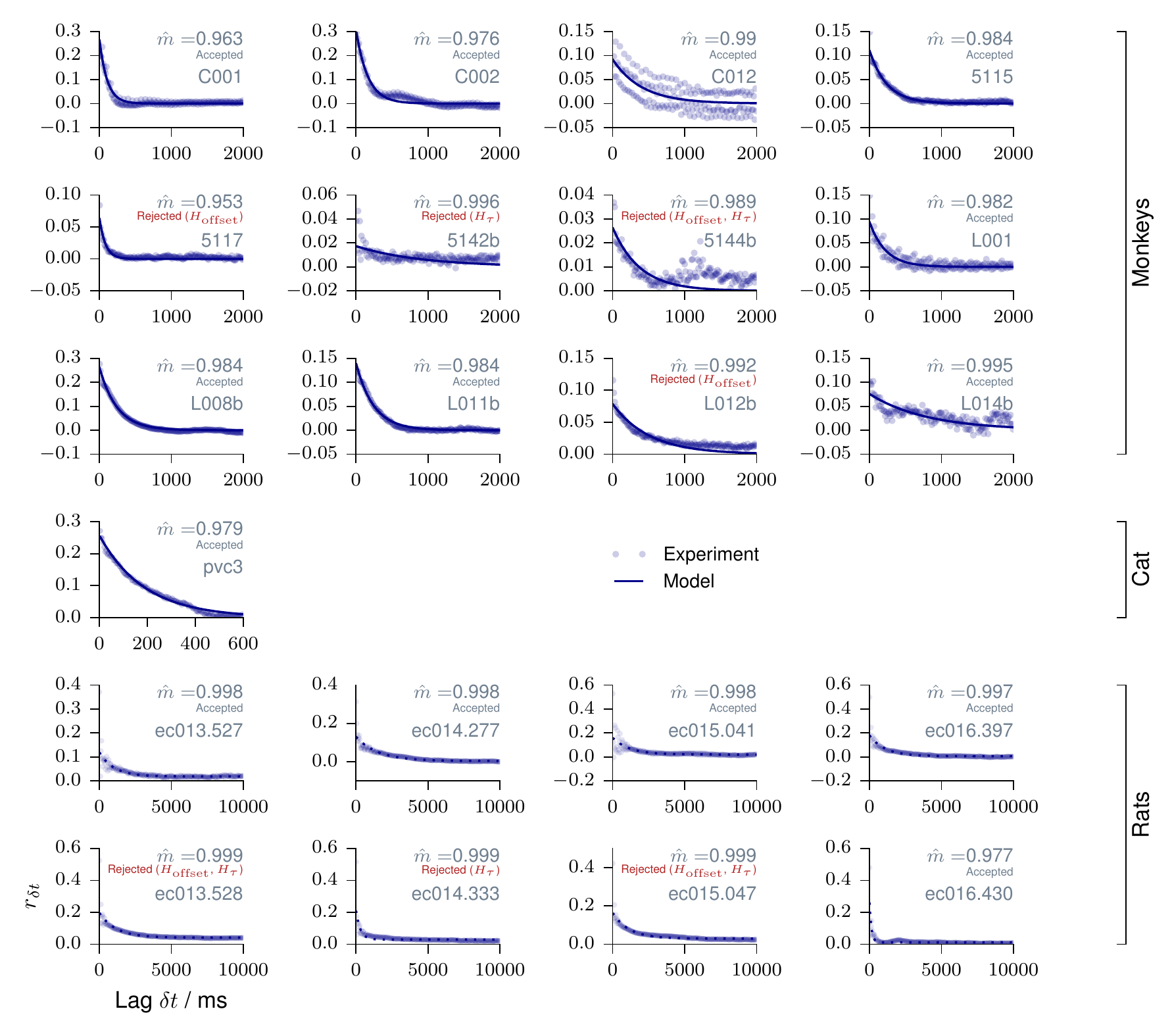}
\caption{\textbf{MR estimation for individual recording sessions.} Reproduced from \cite{Wilting2018}. MR estimation is shown for every individual animal. The consistency checks are detailed in \cite{Wilting2018}. Data from monkey were recorded in prefrontal cortex during an working memory task. The third panel shows a oscillation of $r_k$ with a frequency of 50 Hz, corresponding to measurement corruption due to power supply frequency. Data from anesthetized cat were recorded in primary visual cortex. Data from rat were recorded in hippocampus during a foraging task. In addition to a slow exponential decay, the slopes $r_k$ show the $\vartheta$-oscillations of 6 -- 10 Hz present in hippocampus.}
\label{fig:supp_animal_data}
\end{figure}

\change{In order to test for the applicability of a MR estimation, we used a set of conservative tests \autocite{Wilting2018}, which found the expected exponential relation \change{$r_{\delta t} = b \, m^{\delta t / \Delta t}$} in the majority of experimental recordings (14 out of 21, Fig. \ref{fig:supp_animal_data}).}

\paragraph*{Avalanche size distributions.}
Avalanche sizes were determined similarly to the procedure described in \cite{Priesemann2009,Priesemann2014}. Assuming that individual avalanches are separated in time, let $\lbrace t_i \rbrace$ indicate bins without activity, $a_{t_i} = 0$. The size $s_i$ of one avalanche is defined by the integrated activity between two subsequent bins with zero activity:
\begin{equation}
s_i = \sum_{t = t_i}^{t_{i+1}} a_t.
\end{equation}
From the sample $\lbrace s_i \rbrace$ of avalanche sizes, avalanche size distributions $p(s)$ were determined using frequency counts.
For illustration, we applied logarithmic binning, i.e. exponentially increasing bin widths for $s$.

For each experiments, these empirical avalanche size distributions were compared to avalanche size distributions obtained in a similar fashion from three different matched models (see below for details). Model likelihoods $\mathcal{l}(\lbrace s_i \rbrace) \cond m)$ for all three models were calculated following \cite{Clauset2009}, and we considered the likelihood ratio to determine the most likely model based on the observed data.

\paragraph*{ISI distributions, Fano factors and spike count cross-correlations.}
For each experiment and corresponding reverberating branching model (subsampled to a single unit), ISI distributions were estimated by frequency counts of the differences between subsequent spike times for each channel.

We calculated the single unit Fano factor $F=\Var{a_t} / \E{a_t}$ for the binned activity $a_t$ of each single unit, with the bin sizes indicated in the respective figures.
Likewise, single unit Fano factors for the reverberating branching models were calculated from the subsampled and binned time series.

From the binned single unit activities $a^1_t$ and $a^2_t$ of two units, we estimated the spike count cross correlation $r_\mathrm{sc} = \mathrm{Cov}(a^1_t, a^2_t) / \sigma_{a^1_t}\sigma_{a^2_t}$. The two samples $a^1_t$ and $a^2_t$ for the reverberating branching models were obtained by sampling two randomly chosen neurons.

\subsection{Branching processes}
\label{sec:supp_bps}
In a branching process (BP) with immigration \autocite{Harris1963,Heathcote1965,Pakes1971} each unit $i$ produces a random number $y_{t,i}$ of units in the subsequent time step.
Additionally, in each time step a random number $h_t$ of units immigrates into the system (drive).  
Mathematically, BPs are defined as follows \autocite{Harris1963, Heathcote1965}: 
Let $y_{t,i}$ be independently and identically distributed non-negative integer-valued random variables following a law $Y$ with mean $m = \E{Y}$ and variance $\sigma^2 = \Var{Y}$.
Further, $Y$ shall be non-trivial, meaning it satisfies $\Prob{Y = 0} > 0$ and $\Prob{Y = 0} + \Prob{Y = 1} < 1$.
Likewise, let $h_t$ be independently and identically distributed non-negative integer-valued random variables following a law $H$ with mean rate $h = \E{H}$ and variance $\xi^2 = \Var{H}$.
Then the evolution of the BP $A_{t}$ is given recursively by

\begin{equation}
A_{t+1} = \sum_{i = 1}^{A_{t}} y_{t,i} + h_t,
\label{eq:supp_branching_process}
\end{equation}
i.e. the number of units in the next generation is given by the offspring of all present units and those that were introduced to the system from outside.

The stability of BPs is solely governed by the mean offspring $m$. In the subcritical state, $m<1$, the population converges to a stationary distribution $A_\infty$ with mean $\E{A_\infty} = h / (1 - m)$.
At criticality ($m=1$), $A_{t}$ asymptotically exhibits linear growth, while in the supercritical state ($m>1$) it grows exponentially.

We will now derive results for the mean, variance, and Fano factor of subcritical branching processes.
Following previous results, taking expectation values of both sides of Eq. \eqref{eq:supp_branching_process} yields $\E{A_{t+1}} = m \E{A_{t}} + h$. Because of stationarity $\E{A_{t+1}} = \E{A_{t}} = \E{A_\infty}$ and the mean activity is given by
\begin{align}
\E{A_\infty} = \frac{h}{1-m}.
\label{eq:supp_bp_mean}
\end{align}

\noindent
In order to derive an expression for the variance of the stationary distribution, observe that by the theorem of total variance, $\Var{A_{t+1}} = \E{\Var{A_{t+1} \cond A_{t}}} + \Var{\E{A_{t+1} \cond A_{t}}}$, where $\E{\cdot}$ denotes the expected value, and $A_{t+1} \cond A_{t}$ conditioning the random variable $A_{t + 1}$ on $A_{t}$.
Because $A_{t+1}$ is the sum of independent random variables, the variances also sum: $ \Var{A_{t+1} \cond A_{t}} = \sigma ^2 \, A_{t} + \xi^2$.
Using the previous result for $\E{A_\infty}$ one then obtains

\begin{align*}
\Var{A_{t+1}} = \xi^2 + \sigma ^2 \frac{h}{1-m} + \Var{ m A_{t} + h} =  \xi^2 + \sigma ^2 \frac{h}{1-m} + m^2 \Var{A_{t}}.
\end{align*}

\noindent
Again, in the stationary distribution $\Var{A_{t+1}} = \Var{A_{t}} = \Var{A_\infty}$ which yields

\begin{align}
\Var{A_\infty} = \frac{1}{1-m^2} \left( \xi^2 + \sigma^2 \frac{h}{1-m} \right),
\label{eq:supp_bp_var}
\end{align}

\noindent
The Fano factor $F_{A_{t}} = \Var{A_{t}} \, / \, \E{A_{t}}$ is easily computed from \eqref{eq:supp_bp_mean} and \eqref{eq:supp_bp_var}:

\begin{align}
F_{A_{t}} = \frac{\xi^2}{h (1 +m)} + \frac{\sigma^2}{1 - m^2}.
\label{eq:fano_factor}
\end{align}

\noindent
Interestingly, the mean rate, variance, and Fano factor all diverge when approaching criticality (given a constant input rate $h$): $\E{A_\infty}\rightarrow \infty$, $\qquad \Var{A_\infty}\rightarrow \infty$, and  $F_{A_{t}} \rightarrow \infty$ as $m \rightarrow 1$.

These results were derived without assuming any particular law for $Y$ or $H$. Although the limiting behavior of BPs does not depend on it \autocite{Harris1963,Heathcote1965,Pakes1971}, fixing particular laws allows to  simplify these expressions further.

We here chose Poisson distributions with means $m$ and $h$ for $Y$ and $H$ respectively: $y_{t,i} \sim \mathrm{Poi}(m)$ and $h_t \sim \mathrm{Poi}(h)$.
We chose these laws for two reasons:
(1) Poisson distributions allow for non-trivial offspring distributions with easy control of the branching ratio $m$ by only one parameter.
(2) For the brain, one might assume that each neuron is connected to $k$ postsynaptic neurons, each of which is excited with probability $p$, motivating a binomial offspring distribution with mean $m = k \, p$.
As in cortex $k$ is typically large and $p$ is typically small, the Poisson limit is a reasonable approximation.
Choosing these distributions, the variance and Fano factor become
\begin{align}
\Var{A_{t}} = & \: h \, / \, ((1-m)^2(1+m)), \nonumber \\
F_{A_{t}} = &\: 1 \, / \, (1 - m^2).
\label{eq:supp_bp_simple_var_fano}
\end{align}
Both diverge when approaching criticality ($m=1$).

\subsection{Subsampling}
\label{sec:supp_subsampling}
A general notion of subsampling was introduced in \citet{Wilting2018}.
The subsampled time series $a_{t}$ is constructed from the full process $A_{t}$ based on the three assumptions: 
(i) The sampling process does not interfere with itself, and does not change over time. Hence the realization of a subsample at one time does not influence the realization of a subsample at another time, and the conditional \textit{distribution} of $(a_{t}|A_{t})$ is the same as $(a_{t^\prime}|A_{t^\prime})$ if $A_{t} = A_{t^\prime}$. However, even if $A_{t} = A_{t^\prime}$, the subsampled $a_{t}$ and $a_{t'}$ do not necessarily take the same value.
(ii) The subsampling does not interfere with the evolution of $A_{t}$, i.e. the process evolves independent of the sampling.
(iii) \textit{On average} $a_{t}$ is proportional to $A_{t}$ up to a constant term, $\E{a_t \cond A_t} = \alpha A_t + \beta$. 

In the spike recordings analyzed in this study, the states of a subset of neurons are observed by placing electrodes that record the activity of the same set of neurons over the entire recording.
This implementation of subsampling translates to the general definition in the following manner: 
If $n$ out of all $N$ neurons are sampled, the probability to sample $a_{t}$ active neurons out of the actual $A_{t}$ active neurons follows a hypergeometric distribution, $a_{t} \sim \mathrm{Hyp}(N, n, A_{t})$.
As $\E{a_{t} \cond A_{t} = j} = j \, n \, / \, N$, this representation satisfies the mathematical definition of subsampling with $\alpha = n \, / \, N$.
Choosing this special implementation of subsampling allows to derive predictions for the Fano factor under subsampling and the spike count cross correlation.
First, evaluate $\Var{a_{t}}$ further in terms of $A_{t}$:

\begin{align}
\Var{a_{t}} = & \: \E{\Var{a_{t} \cond A_{t}}} + \Var{\E{a_{t} \cond A_{t}}} \nonumber \\
= & \: n \E{ \frac{A_{t}}{N}\frac{N - A_{t}}{N} \frac{N-n}{N-1} } + \Var{\frac{n}{N}A_{t}} \nonumber \\
= & \: \frac{1}{N}\frac{n}{N}\frac{N-n}{N-1} \left( N \, \E{A_{t}} - \E{A_{t}^2} \right) + \frac{n^2}{N^2} \Var{A_{t}} \nonumber \\
= & \: \frac{n}{N^2}\frac{N-n}{N-1} \left( N \, \E{A_{t}} - \E{A_{t}}^2 \right) + \left( \frac{n^2}{N^2} -  \frac{n}{N^2}\frac{N-n}{N-1} \right) \Var{A_{t}}.
\label{eq:subsampled_variance}
\end{align}

\noindent
This expression precisely determines the variance $\Var{a_{t}}$ under subsampling from the properties $\E{A_{t}}$ and $\Var{A_{t}}$ of the full process, and from the parameters of subsampling $n$ and $N$.
We now show that the Fano factor approaches and even falls below unity under strong subsampling, regardless of the underlying dynamical state $m$. In the limit of strong subsampling ($n \ll N$) Eq. \eqref{eq:subsampled_variance} yields:

\begin{align}
\Var{a_{t}} \approx \frac{n}{N^2} \left( N \E{A_{t}} - \E{A_{t}}^2 \right) + \frac{n^2 - n}{N^2} \Var{A_{t}}.
\label{eq:supp_hypergeom_subsampled_variance}
\end{align}

\noindent
Hence the subsampled Fano factor is given by

\begin{align}
F_{a_{t}} = \frac{\Var{a_{t}}}{\E{a_{t}}} \approx 1 - \frac{\E{A_{t}}}{N} + \frac{n-1}{N} \frac{\Var{A_{t}}}{\E{A_{t}}} = 1 - \frac{\E{A_{t}} - (n-1)F_{A_{t}}}{N}.
\label{eq:supp_hypergeom_subsampled_fano}
\end{align}

\noindent
Interestingly, when sampling a single unit ($n=1$) the Fano factor of that unit becomes completely independent of the Fano factor of the full process:

\begin{equation}
F_{a_{t}} = 1 - \E{A_{t}} / N = 1 - \E{a_{t}} / n = 1 - R,
\end{equation}

\noindent
 where $R = \E{a_{t}} / n$ is the mean rate of a single unit.

Based on this implementation of subsampling, we derived analytical results for the cross-correlation between the activity of two units on the time scale of one time step.
The pair of units is here represented by two independent samplings $a_{t}$ and $\tilde{a}(t)$ of a BP $A_{t}$ with $n = 1$, i.e. each represents one single unit.
Because both samplings are drawn from identical distributions, their variances are identical and hence the correlation coefficient is given by $r_\mathrm{sc} = \mathrm{Cov}(a_{t}, \tilde{a}(t)) \, / \Var{a_{t}}$. Employing again the law of total expectation and using the independence of the two samplings, this can be evaluated:

\begin{align}
\mathrm{Cov}(a_{t}, \tilde{a}(t)) =  \E[A_{t}]{ \E{a_{t} \, \tilde{a}(t) \cond A_{t}} } -  \E[A_{t}]{ \E{a_{t} \cond A_{t}}}^2 = \frac{1}{N^2}	\Var{A_{t}},
\end{align}

\noindent
with the first inner expectation being taken over the joint distribution of $a_{t}$ and $\tilde{a}(t)$.
Using Eq. \eqref{eq:supp_hypergeom_subsampled_variance}, one easily obtains
\begin{equation}
r_\mathrm{sc} = \frac{\Var{A_{t}}}{N \E{A_{t}} - \E{A_{t}}^2} = \frac{F_{A_{t}}}{N - \E{A_{t}}} = \frac{F_{A_{t}}}{N \, (1  - R)}
\label{eq:supp_spike_count_correlation}
\end{equation} 

\noindent
with the mean single unit rate $R = \E{A_{t}} / N$. For subcritical systems, the Fano factor $F_{A_{t}}$ is much smaller than $N$, and the rate is typically much smaller than 1. Therefore, the cross-correlation between single units is typically very small.



\newpage

\newpage

\newpage

\newpage

\newpage

\newpage

\newpage

\newpage

\newpage

\newpage

\newpage

\end{document}